\documentclass[twocolumn]{aastex63}

\shorttitle{CWITools}
\shortauthors{O'Sullivan and Chen}
\graphicspath{{./}{figures/}}

\begin{document}

\title{CWITools: A Python3 Data Analysis Pipeline for the Cosmic Web Imager Instruments}

\correspondingauthor{Donal O'Sullivan}
\email{dosulliv@caltech.edu, donal.b.osullivan@gmail.com}

\author{Donal O'Sullivan}
\affiliation{California Institute of Technology}

\author{Yuguang Chen}
\affiliation{California Institute of Technology}

\begin{abstract}
The Palomar Cosmic Web Imager (PCWI) and Keck Cosmic Web Imager (KCWI) are integral-field spectrographs on the Hale 5m telescope at Palomar Observatory and the Keck-2 10m telescope at W. M. Keck Observatory, respectively. In recent years, these instruments have been increasingly used to conduct survey work; in particular focused on the circumgalactic and intergalactic media at high redshift. Extracting faint signals from three-dimensional IFU data is a complex task which can become prohibitively difficult for large samples without the proper tools. We present CWITools, a package written in Python3 for the analysis of PCWI and KCWI data. CWITools is designed to provide a pipeline between the output of the standard instrument data reduction pipelines and scientific products such as surface brightness maps, spectra, velocity maps, as well as a wide array of associated models and measurements. While the package is designed specifically for PCWI and KCWI data, the package is open source and can be adapted to accommodate any three-dimensional integral field spectroscopy data. Here, we describe this pipeline, the methodology behind individual steps and provide example applications.\\
\end{abstract}

\keywords{Python3, Data Analysis, Cosmic Web Imager, PCWI, KCWI}

\section{Introduction}
The Palomar and Keck Cosmic Web Imagers (hereafter \emph{KCWI} and \emph{PCWI}) are integral field unit (IFU) spectrographs designed to study faint, extended emission \citep{Matuszewski:2010,Morrissey:2018}. PCWI was installed on the Hale 5m telescope at Palomar Observatory in 2009, while KCWI was installed on the Keck-2 10m telescope in the W. M. Keck Observatory in 2017. In 2014, the Multi-unit Spectroscopic Explorer (MUSE) \citep{Caillier:2014} was installed on the 8m VLT at the European Southern Observatory. This new set of instruments on 5-10m class telescopes has enabled observers to directly detect signals on the order of $10^{-18}$ erg/s/cm$^2$/arcsec$^2$ in less than an hour of telescope time \citep{Martin:2014a,Martin:2014b}. This in turn has enabled surveys of unprecedented sizes mapping the circumgalactic medium around high-redshift galaxies and quasars \citep{Borisova:2016,ArrigoniBattaia:2019a,OSullivan:2020a,Cai:2019}. As the observational field grows and sample sizes increase, data analysis becomes an increasingly prevalent issue. Here we present CWITools, a data analysis Python3 toolkit tailored to handling the data produced by the PCWI and KCWI data reduction pipelines (DRPs). This toolkit can be seen and used as a pipeline itself, taking input in the form of individual data cubes and producing scientific products such as white-light images, pseudo-narrow-band images, spectra, and velocity maps. CWITools was intially built out of necessity, as a toolkit for work on the FLASHES Survey \citep{OSullivan:2020a}; a survey of extended emission in $z=2-3$ QSO environments. Over the past two years, it has been transformed into a publicly available, user-friendly package with help menus, documentation and application examples. This package is open source, and can be adapted to work with any three-dimensional data. However, in order to limit the scope for the purposes of testing and robustness, we focus primarily on applications involving data from PCWI and KCWI.\\

We begin by providing an overview of the context and architecture of CWITools, including a detailed description of the PCWI and KCWI pipelines' output. We then describe the methodology of each broad processing step within CWITools; reduction, extraction, synthesis, modeling, and measurement. Finally, we provide an example application of CWITools in detecting nebular emission around a source at high redshift. In general, since full code-specific documentation and examples exist online for the package, we will limit the discussion here to design, methodology, application examples, and recommendations.

\section{Package Architecture} 

\begin{figure*}[t]
    \centering
    \includegraphics[width=\textwidth]{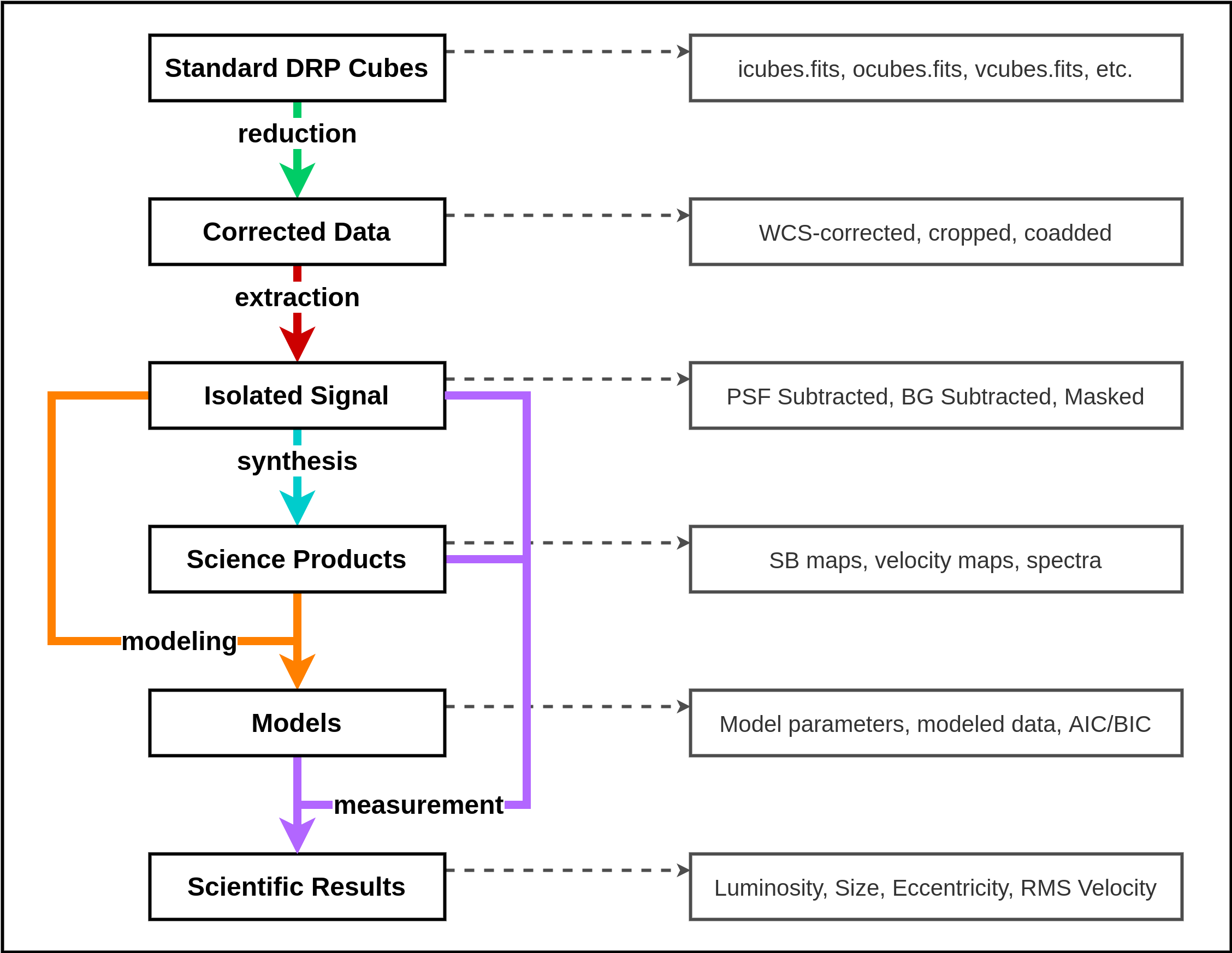}
    \caption{The architecture of CWITools, showing the broad pipeline from standard DRP results to scientific results, as well as the nature of the input and output of each module. Arrows represent different modules within the package and boxes represent the different types of data. For example, at the top, the reduction module (green arrow) takes standard DRP data cubes as input, and outputs corrected/coadded data cubes. On the right hand side are specific examples of the kinds of data products at each stage.}
    \label{CWITools:figure:architecture}
\end{figure*}

CWITools is intended to provide a bridge between the output of the standard instrument data reduction pipelines (DRPs) and scientific products. In particular, PCWI and KCWI are both designed to image faint and diffuse emission, which lends a particular scientific focus to the package, though by no means an exclusive one. While there is no ubiquitous procedure which applies to all scientific projects, there are certain steps which are more or less universal to the process of extracting and measuring signals in IFU data: cropping, correcting 3D coordinate systems, masking or subtracting sources, extracting spectra or velocity maps, etc. CWITools is thus intended to provide observers with a flexible and easy-to-use set of tools with which they can customize a pipeline to suit their needs.\\ 

The standard DRPs have the goal of producing fully calibrated, three-dimensional data cubes for each individual exposure. Typically, observers reduce data on a night-by-night basis. Others may organize their data based on instrument configurations used or by target observed during a given observing run. For the purpose of scientific work, it makes the most sense to organize any subsequent analysis by science target. As such, the central element of any CWITools is a `.list' file which indicates simply: (i) where the input data for this target can be found, (ii) a set of unique IDs for files associated with this target, and (iii) where to save output and intermediate products. \\ 

The end point of CWITools is whatever scientific measurement is needed for the discussion and analysis part of a project. Whatever the scientific case - the functional architecture can be broken into the same main modular components: reduction, extraction, synthesis, modeling, and measurement. Reduction involves steps which are in essence further corrections of the data, required to compile the final observation (e.g. cropping, coadding). Extraction refers to steps which are focused on isolating a specific element or signal within the data (e.g. removing foreground or background). Synthesis is the generation of first-level scientific products from the isolated signal, such as surface brightness maps or velocity maps. Modeling is the fitting and evaluation of models, applied either to the generated scientific products or directly to the isolated 3D signal. Finally, measurement is the calculation of physical quantities from synthesized products, 3D data, or models. The distinction between \textit{synthesis} and \textit{measurement} in the context of this package is that the former produces 1D-3D data structures (e.g. spectra or velocity maps) upon which further measurements or analysis can be performed, while the latter produces scalar results which represent end-points in the data analysis process (e.g. luminosity). Although the term `pipeline' implies a waterfall-like process through the above steps and the bulk of pipelines will go through the above steps in some similar order, there is no strict one-directional flow imposed by the design of the package. There is always a point beyond which automation becomes more cumbersome than the alternative manual work required and, in the context of this package, that point is reached in determining the exact order of operations. Steps often need to be skipped, re-arranged, or repeated depending on the scientific objective and there is no `one size fits all' data analysis pipeline. As such, while CWITools provides a number of template pipelines as examples, they are intended as templates to be modified and adapted to observers' needs.\\ 

Figure~\ref{CWITools:figure:architecture} shows the internal modular structure of CWITools, as well as the associated inputs and outputs of each stage. Each module corresponds to a Python module within the main package (e.g. \texttt{cwitools.reduction}), within which are functions associated with that stage. In addition to these modules, which can be imported into Python environments and used to construct a pipeline within any scripting environment, CWITools contains a library of command-line scripts which serve as wrappers to these functions. These scripts are designed as an interface for users who are less familiar with Python scripting, and simply want access to the tools. Upon installation, they are added to the user's environment as terminal commands (e.g. `\texttt{cwi\_coadd}' becomes the direct command to coadd data cubes), each of which has a help menu guiding the user on its usage. Users who are not familiar with Python can then construct their data analysis pipelines using simple bash scripts containing a number of these commands. Short examples of both a Python environment pipeline and a bash script pipeline are included in the Appendix, and a full set of examples are available within the package data itself. 

\section{Cosmic Web Imager Data Format}
The standard data reduction pipelines of PCWI and KCWI produce three-dimensional data cubes containing two spatial axes and one wavelength axis. In this section, we briefly describe the different output file types produced by the standard DRPs and the 3D coordinate systems of those data structures, so as to lay a foundation for the discussion of the methodology.  

\subsection{Standard DRP Output}
The standard pipelines of both PCWI and KCWI first apply the usual reduction steps - bias correction, flat fielding, dark subtraction, etc. - to the raw 2D detector images. Each 2D image contains 24 2D spectra (one for each slice in the image slicer) arranged side by side. Using a series of calibration images, the pipeline reconstructs these 2D images into 3D data cubes with two spatial axes and one wavelength axis. The first 3D data product is given extension ``cube.fits,'' with each exposure producing a non-sky-subtracted object frame (\textit{ocube}), a sky-subtracted intensity frame (\textit{icube}), a frame of the sky data or sky model used (\textit{scube}), and an associated 3D variance estimate (\textit{vcube}). As the 3D data is refined through subsequent stages, the filenames are updated to reflect the stages. For example, after a slice-to-slice relative-response correction (stage6\_rr in the PCWI DRP), the letter `r' is appended to each so that the files now have the extensions `icube\textbf{r}.fits,' `ocube\textbf{r}.fits,' `scube\textbf{r}.fits,' etc. After flux calibration using a standard star, the appended letter is changed (rather than added) to `s,' so the filenames are now `icube\textbf{s}.fits,' `ocube\textbf{s}.fits,' etc. A full filename will include a unique identifier combined with one of these extensions. For example, the fully reduced, flux-calibrated cube for KCWI exposure number 116 might be `kb200115\_00116\_icubes.fits.' This is of central relevance here because a core operational mode of CWITools involves providing as input a list of unique IDs (e.g. `200115\_00116' for this exposure) and a `cube type' (e.g. icubes.fits) to work with. This allows users a simple interface with a high level of flexibility. For example, if an observer has three exposures for a certain target, they just need to store the three unique IDs for those exposures in a CWITools `.list' file, after which they can run any operation on any data product for that target by providing both the ID list and the desired cube type (e.g. coadd the intensity cubes, then coadd the sky cubes).\\ 

All flux-calibrated CWI data cubes are produced in ``FLAM'' units - i.e. flux per unit wavelength: $F_\lambda \equiv \mathrm{erg~s^{-1}~cm^{-2}~\text{\AA}^{-1}}$. While PCWI outputs directly in $F_\lambda$ units, KCWI data cubes are produced in units of ``FLAM16,'' $F_{\lambda,16}=10^{16}F_\lambda$. Non-flux-calibrated data cubes have units of `electrons' - i.e. the number of photo-electrons measured in each voxel.

\subsection{Coordinate Systems and Headers}
There are three coordinate systems which are of relevance when analyzing CWI data (and IFU data in general). The first is the \emph{world coordinate system} (WCS) - which refers to the real world measurements of wavelength ($\lambda$), right-ascension ($\alpha$), and declination ($\delta$). The second is the \emph{image coordinate system}, referring to the axes within the data cube. Let us denote these as $x$, $y$, and $z$, where $x$ and $y$ are spatial axes and $z$ is the wavelength axis. It is important to note that the world-coordinate axes $\alpha$ and $\delta$ only correspond directly to $x$ and $y$ when (i) the position angle is a multiple of $90^{\circ}$ and (ii) the field of view is small. In general, a one-to-one correspondence between image and WCS axes should not be taken for granted; i.e. $\alpha \rightarrow \alpha(x,y)$ and $\delta \rightarrow \delta(x,y)$. \\

FITS image formats contain ``header'' objects which store meta-data about the image such as timestamps, configuration details, and exposure times. The headers also contain the necessary information to translate between the two above coordinate systems. Specifically, they contain sets of keywords to define (i) the number of axes and size of each axis, (ii) a reference point in the image for a known world coordinate (e.g. the right-ascension and declination at a given $x,y$), and (iii) the change in world coordinates along each image axis (e.g. the change in $\alpha$ along the $x$ axis). Table~\ref{CWITools:Table:CDMatrix} lists these keywords and their meanings. The widely used package Astropy provides a convenient way to handle coordinate systems by creating WCS objects which store this information and provide some useful functions such as mapping $(x, y, z)$ to $(\alpha, \delta, \lambda)$ and vice versa \citep{AstropyI,AstropyII}.\\

\begin{deluxetable*}{lll}
\caption{A summary of FITS Header keywords for world coordinate systems. \label{CWITools:Table:CDMatrix}}

\tablehead{ \colhead{Keyword} & \colhead{Description} & \colhead{Example} }
\startdata
NAXIS           & The number of axes                            & 3\\
NAXIS\textbf{A} & The length of image/data axis \textbf{A}               & 127 \\ \hline
CTYPE\textbf{B} & The type of world coordinate axis \textbf{B}  & RA--TAN \\
CNAME\textbf{B} & The name of world coordinate axis \textbf{B}  & KCWI RA \\
CUNIT\textbf{B} & The units of world coordinate axis \textbf{B} & deg \\ 
CRVAL\textbf{B} & The reference value for world coordinate axis \textbf{B} & 255.25857 \\
CRPIX\textbf{B} & The reference pixel along image axis \textbf{B} & 32 \\ \hline 
CD\textbf{A}\_\textbf{B} & Change in world axis \textbf{B} per pixel of image axis \textbf{A}& $8.09\times10^{-5}$  \\
\enddata
\end{deluxetable*}

As a final note on coordinate systems, a common source of confusion is the varying conventions when it comes to ordering axes and defining origins. The FITS headers for CWI data specify the axes such that the order of the axes is $(1, 2, 3) = (x, y, w)$. However, when loading the data in a Python shell (e.g. with AstroPy or NumPy), the data structure has the order of axes reversed: $(1, 2, 3) = (w, y, x)$. Furthermore, while the values in FITS headers are 1-indexed (i.e. the index of the first pixel is 1), data structures in Python are typically 0-indexed. This must be taken into account when handling header keywords such as CRPIX1, or converting between coordinate systems. 

\begin{figure*}[t]
    \centering
    \includegraphics[width=0.95\textwidth]{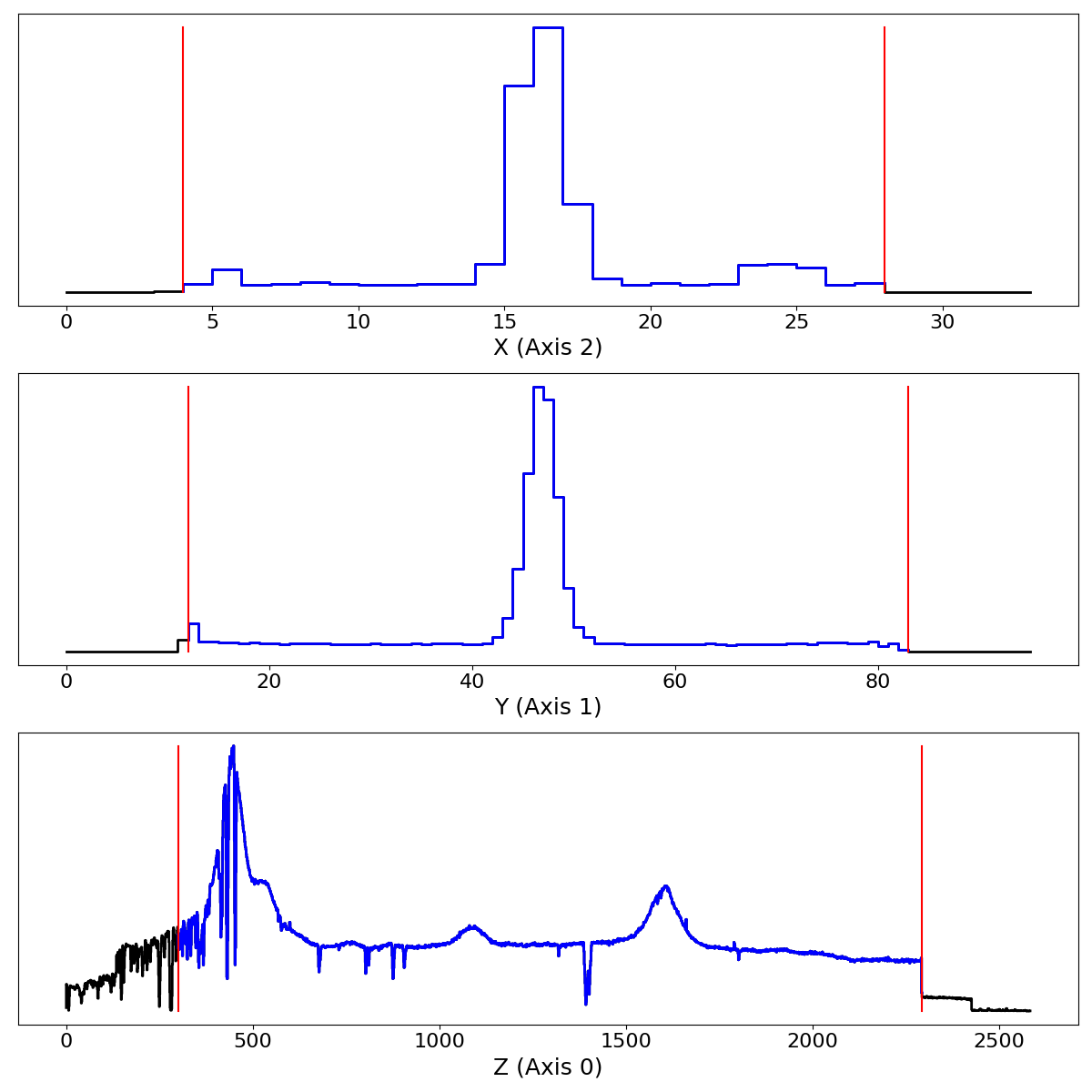}
    \caption{Automatic cropping parameters obtained by CWITools. This view is presented to the user if automatic cropping is requested. Each panel presents a summed one-dimensional profile for a different axis ($x$, $y$, $z$ from top to bottom). Data within the cropped range, delineated by vertical red lines, is highlighted in blue, while data outside the range is black. This is most useful as a first step, from which the user can determine the best cropping parameters to suit their needs.}
    \label{CWITools:Figure:AutoCrop}
\end{figure*}

\section{Module: Reduction} 

The reduction module is focused on steps for further refinement of the final observational data. This includes any steps beyond the standard date reduction pipeline which are required to create the final, fully calibrated, combined data cube for a given target. In this section, we will describe each of these steps in detail.

\subsection{Cropping}
The output data cubes from both the PCWI and KCWI DRPs both require cropping along all three axes. While a user can determine the crop parameters they want to use, there are some defaults determined by the nature of the final PCWI and KCWI data cubes. The layout of the 2D spectra of the slices in detector space is such that alternating slices are staggered in wavelength. This means that the bandpass for each slice is slightly different, and the wavelength range which is common to all slices is slightly less than the instantaneous bandpass of any one slice. Only data within this common overlap region can be reliably calibrated by the DRP. The headers of KCWI and PCWI FITS files contain the keywords `WAVGOOD0' and `WAVGOOD1' which indicate this range. Therefore, as a default, CWITools will crop the wavelength to this range.\\ 
\begin{figure*}[t]
    \centering
    \includegraphics[width=0.95\textwidth]{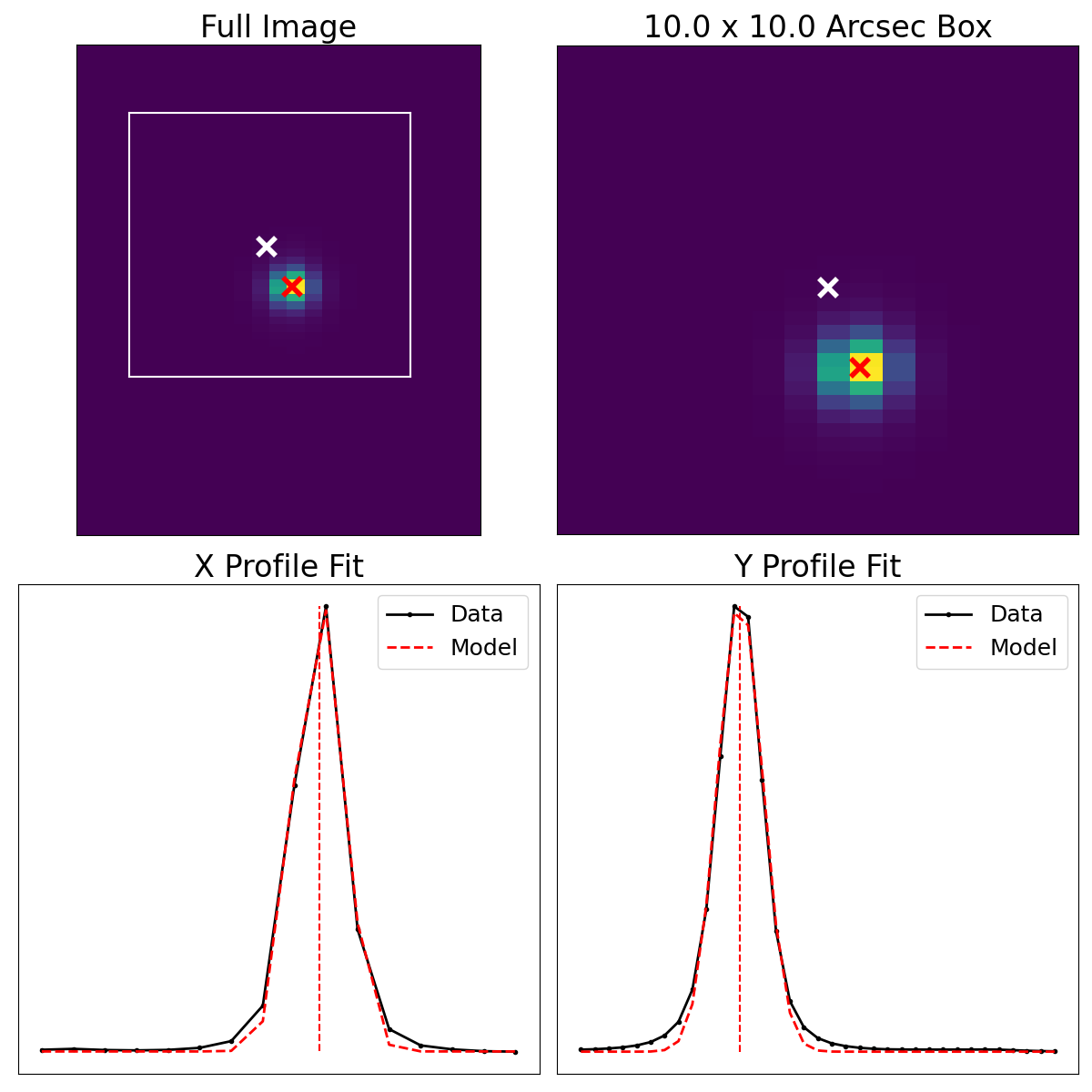}
    \caption{Automatic WCS correction using source fitting. As one option for spatial WCS correction, CWITools assumes that the initial WCS is approximately correct, then identifies and fits the nearest source to that location. The above view is the view presented to the user during this step if requested, so that the user can inspect the fit visually. The top left panel shows the full field of view, with a white cross indicating the expected location of the primary source. The white box indicates the search area, which can be adjusted by the user. The red cross indicates the fitted location of the source. The top right panel shows a zoom in on the white box. The bottom two panels show the one-dimensional PSF of the source along each axis and a simple 1D Gaussian fit to the data.}
    \label{CWITools:Figure:AutoWCS}
\end{figure*}

Spatially, there are different reasons to crop PCWI and KCWI data. For PCWI, the x-axis (FITS axis 1, NumPy axis 2) of the data contains some buffer, going slightly beyond the edge of each slice. The exact margin can be determined by looking at a fully reduced cube from a continuum flat image, but is usually approximately about 10 pixels on either side. There is no padding or margin along the y-axis, which contains only the 24 slices. For fully reduced KCWI data, there is padding if stage 7 (differential atmospheric refraction correction) has been applied, and the amount of padding differs depending on the slicer setting, so there is no hard-coded default for the spatial padding. Instead, the `auto-crop' mode (i.e. the default used in absence of user input) is to trim empty rows and columns. Users should be aware that this may not be sufficient to avoid edge artifacts which may be present in the data. The cropping tool has a plot functionality which can be used to view the crop settings overlaid on a profile of each axis. This is a helpful tool in selecting the best settings. As with all steps that modify existing cubes, the output is saved by default with a modified file extension. In this case, the ``.fits'' of the input is replaced with ``.c.fits,'' to indicate that it has been cropped. 

\begin{figure*}[t]
    \centering
    \includegraphics[width=\textwidth]{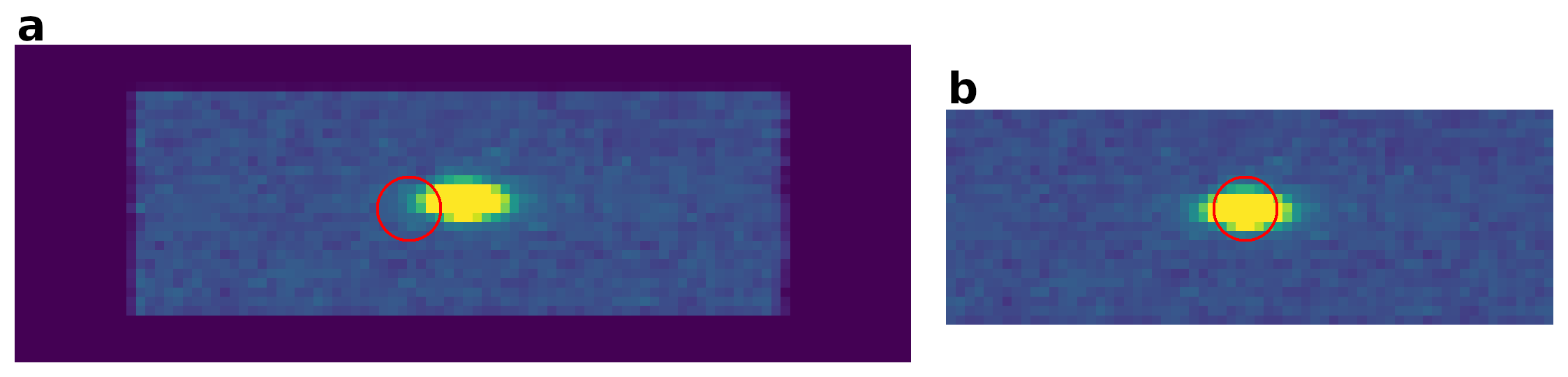}
    \caption{Cropping and WCS correction applied to an individual cube. Left (a): a spatial 2D snapshot of the data cube prior to cropping and WCS correction. The red circle has the correct coordinates for the source, SDSS0958+4703. Right (b): the cropped and WCS-corrected cube, the red circle now aligns with the visible source.}
    \label{CWITools:Figure:WCS_Crop}
\end{figure*}

\subsection{World Coordinate System Correction} 

As described earlier, the world coordinate system (WCS) is the three-dimensional coordinate system of right ascension ($\alpha$), declination ($\delta$), and wavelength ($\lambda$). FITS headers for three-dimensional data contain reserved key words which determine the translation from image coordinates ($x$, $y$, $z$) to world coordinates ($\alpha$, $\delta$, $\lambda$). To constrain the transformation, several pieces of information are provided. The first is the position of a specified world coordinate in image coordinates; i.e. that the 3D position ($\alpha_0$, $\delta_0$, $\lambda_0$) is coincident with ($x_0$, $y_0$, $z_0$). The values $\alpha_0$, $\delta_0$, and $\lambda_0$ are stored in `CRVAL' keywords, where the central value for each specified world-coordinate axis is given. Since the order of the world coordinate axes is usually right-ascension, declination, and wavelength, this means the header contains `CRVAL1=$\alpha_0$,' `CRVAL2=$\delta_0$,' and `CRVAL3=$\lambda_0$.' It is important to clarify that the numbering here is for the values associated  with the world-coordinate axes, not with image axes. The coincident image coordinate is stored in `CRPIX' keywords: `CRPIX1=$x_0$,' `CRPIX2=$y_0$' and `CRPIX3=$z_0$'. Finally, the change in each world coordinate axis along each image is provided by the ``CD Matrix,'' which is a set of keywords of the form `CD\textbf{W}\_\textbf{I},' which specifies the change in world-coordinate axis \textbf{W} per pixel along image axis \textbf{I}. As an example, `CD1\_2' encodes the change in right-ascension per pixel along the $y$-axis. The units of these values are given by keywords `CUNIT1,' `CUNIT2,' and `CUNIT3.' \\ 

Generally, the CD Matrix can be taken as accurate for all PCWI and KCWI data cubes. The only rectification that is usually required is an adjustment of the central reference point. This is done separately for the spatial and wavelength axes. In each case, there are two basic approaches to choose from: measuring the location of a feature with known world-coordinates or cross-correlating the input data so that they are at least aligned. The former provides a correction in absolute terms, but requires a measurable source with known coordinates, which is not always available. The latter provides a fall-back for these cases, such that the input data can be ensured to have consistent world-coordinate systems, but the absolute values may remain inaccurate. Each of the four processes is described below. Cubes with corrected coordinates systems are saved by default with the added file extension ``.wc.fits'' (for WCS-corrected).

\subsubsection{Spatial Correction: Source Fitting}
The preferred way to correct the spatial axes is to measure the location of a known source within the image. This is done by first creating a white-light image from the input data (see Synthesis module in Section ~\ref{CWITools:Section:Synthesis}). The default operating assumption is that the initial WCS is approximately correct, and a $10''\times10''$ box around the estimated location of the source is extracted. In the case that the initial WCS is extremely inaccurate, an initial guess of the source location can be provided, and the size of the box can be adjusted. Once the box has been extracted, 1D profiles in $x$ and $y$ are formed by summing along the image axes, and a 1D Moffat profile is fit to the source to obtain the best-fit center. CRPIX1 and CRPIX2 are updated to the $x$ and $y$ centers, respectively, and CRVAL1/CRVAL2 are updated to the known RA/DEC of the source.

\subsection{Wavelength Correction: Line Fitting}
To correct the wavelength axis, a known sky-line can be fit with a simple Gaussian model. The default way to do this is with sky cubes (e.g. ``scubes.fits'') and known sky emission lines. CWITools package data includes a full blue-optical sky spectrum for Keck, and a preliminary set of known emission lines in both the Palomar and Keck blue-optical sky spectra. For example, there is a bright mercury line (thanks to light pollution) in the Palomar night sky, Hg I $\lambda4358.3$. This line was extremely useful in correcting the Palomar data for the FLASHES Pilot Survey~\citep{OSullivan:2020a}. A high-SNR sky spectrum is compiled from the input sky cube by summing over both spatial axes. As in the spatial PSF fitting, the default assumption is that the initial WCS is approximately correct. Therefore, a window of $\Delta \lambda \simeq 10$ \AA~ around the initial estimate of the sky-line is extracted from the spectrum, and a 1D Gaussian model is fitted to the data to obtain the true center. The difference between the initial WCS' estimate of the line position and the fitted position is calculated in units of pixels, and the header keyword CRPIX3 is updated accordingly so that the WCS is consistent with the measured position of the source. 

\subsubsection{Wavelength Correction: 1D Cross-Correlation}

If no spatial source or sky line is available, cross-correlation can be used to ensure that the input data are all self-consistent and aligned, even if the absolute world-coordinate solution is not exactly known. To do this, as in the 2D cross-correlation above, one image must be picked as the reference point. For each sky cube, a 1D spectrum is then generated and each spectrum is cross-correlated with the reference spectrum. A list of relative offsets, in units of pixels, is then calculated between the spectra. The CRPIX3 header keywords in all but the reference image are updated based on the measured offsets so that they are consistent with the WCS of the reference image. Any error in the reference image's WCS thus remains in the corrected WCS.

\subsubsection{Custom WCS Correction Routines}
Certain science cases, such as creating mosaics of very faint emission, satisfy none of the above requirements and require more advanced, home-made methods. As such, CWITools separates the WCS measurement and WCS correction steps. WCS measurement produces a WCS correction file which contains a table of the desired values for the CRVAL and CRPIX keywords for each input cube. If using either of the in-built methods (feature fitting or cross-correlation), this is automatically generated by the WCS measurement function. If using a more complex method, or in the case that some manual adjustment is required, a user can generate this table themselves by whatever method they desire. As long as the format of the table (which is quite simple) is correct, the table file can be fed as input into the `apply\_wcs()' function. Because CWITools uses the WCS to automatically coadd data, this step is crucial in determining the quality of the final, coadded data. 

\subsection{Coadding} 

\begin{figure*}[t]
    \centering
    \includegraphics[width=\textwidth]{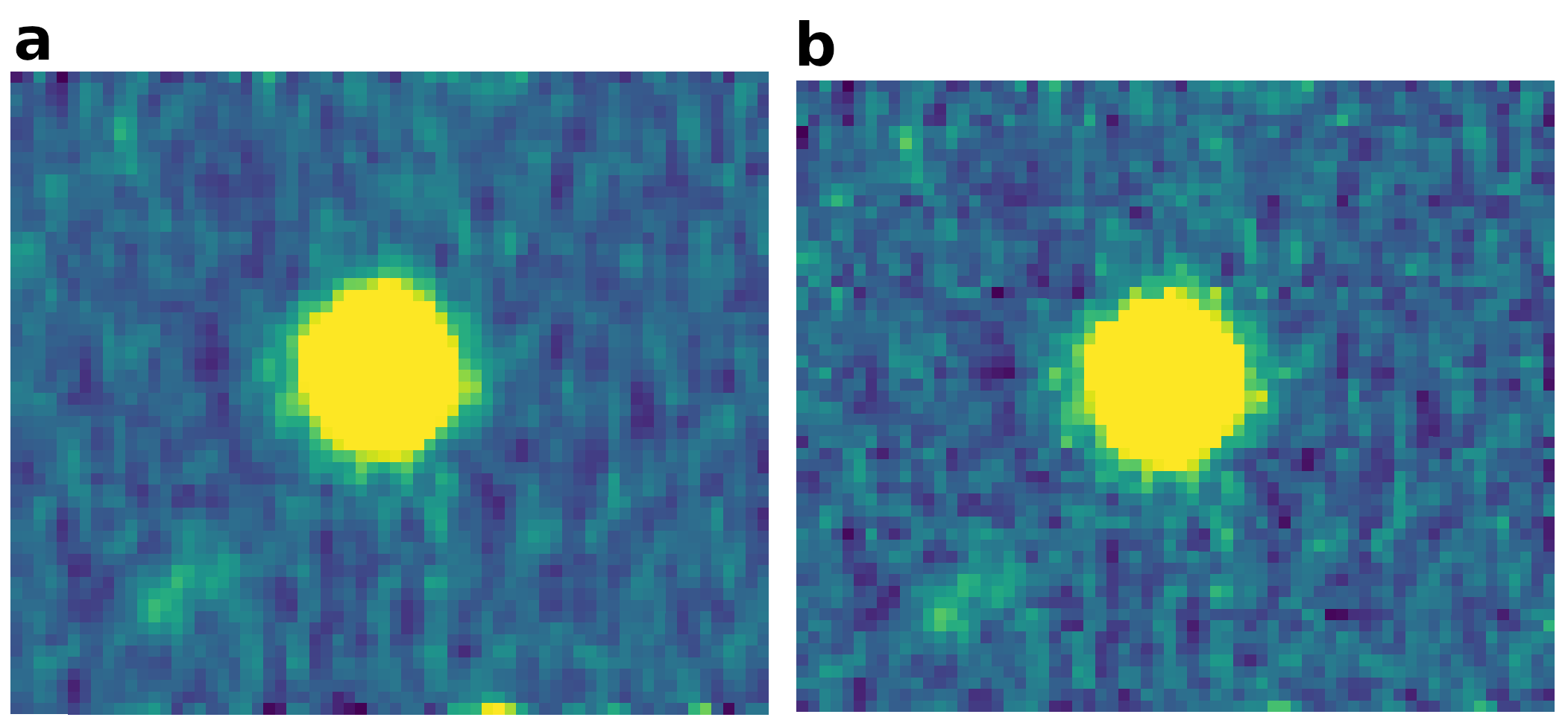}
    \caption{Spatial 2D slices of coadded frames using different drizzle factors. In panel (a) on the left, the coadd was performed normally, with no drizzling (i.e. $f_{drz}=1.0$), while in panel (b), a very low Drizzle factor of $f_{drz}=0.4$ was used. Normally, one would use a factor in the range $f_{drz}\sim 0.6-0.8$, but we use a low factor here to make the visual difference clear. The image in panel (b) is clearly sharper with higher frequency noise, while that in panel (a) looks smooth in comparison.}
    \label{CWITools:Figure:Drizzle}
\end{figure*}

The single most ubiquitous and (relatively speaking) computationally intensive task in PCWI and KCWI data analysis is coadding data onto a common three-dimensional grid, including variance propagation. There is, of course, already a plethora of openly accessible and efficient Python code which performs some subset of this task, but there are several key issues which led us to develop an entirely custom algorithm.\\ 

First, nearly all of the widely used packages implementing coadding or drizzling algorithms (e.g. \cite{Drizzle}) are written with two-dimensional imaging data in mind. Second, knowledge of every computational step is needed in order to accurately propagate variance through the coadd process. This is trivial if the data only needs to be resampled and shifted linearly along its axes, but coadding images with arbitrary rotations makes the resampling - and thus the mathematics of error propagation - significantly more complicated. As such, in order to have an algorithm which can coadd arbitrary input in terms of position angle and spatial sampling, we have developed a custom three-dimensional coadding algorithm, including a drizzling factor. As a final note, CWITool's coadding algorithm makes use of existing 3D mask cubes produced by the PCWI and KCWI pipelines - automatically loading and using them if the user requests it. These cubes flag noisy edge pixels, pixels affected by cosmic rays, and other potentially corrupted pixels. These can then be excluded from the coadd, improving the quality of the final product.\\ 

The CWITools coadd process is split into two main steps. First, the input cubes are aligned in wavelength. At the moment, only input with a common sampling rate in wavelength is accepted, since this is by far the most common scenario, though a future update is planned to allow multiple wavelength sampling rates in the input. The minimum and maximum input wavelengths are determined ($\lambda_{min}$ and $\lambda_{max}$), as well as the input resolution, $\Delta\lambda$. A new common wavelength grid is generated spanning the range $[\lambda_{min}-\Delta\lambda, \lambda_{max}+\Delta\lambda]$ with the same resolution. Then, for each input cube, the cube is padded with zeros along the z-axis until it is the same length as the common wavelength axis, and the offset (in Angstrom) between the first index of the original input grid and the first index of the new common grid is determined. Let $\delta \lambda_{i}$ be this offset. The required shift in pixels is then $\delta z=\delta\lambda_i / \Delta \lambda$. This is split into an integer shift, $\delta z_{INT} = int(\delta z)$, and a sub-pixel shift, $\delta z_{SUB} = \delta z - \delta z_{INT}$. The integer shift component requires no interpolation, and thus no error propagation. The data and associated variance are just rolled along the z-axis by $\delta z_{INT}$. The sub-pixel shift is then performed using linear interpolation, implemented as a convolution with the 1D kernel $K_z = [\delta z_{SUB}, 1 - \delta z_{SUB}]$. To propagate the error on this step, the variance is convolved with $K_z^2$. At the end of this step, the cubes are all aligned in wavelength and have the same z-axis length.\\ 

The second major step is the spaxel-by-spaxel projection of the input cubes' footprints onto a common coadd grid. The on-sky footprints of each input field of view are calculated and the footprint required to encompass all of the input data is derived. The minimum spatial sampling of the input is taken as the uniform spatial sampling of the desired output grid. This information is then used to construct a new header and empty data cube for the coadded data. The 2D (x,y) vertices of each input pixel are mapped from input image coordinates to on-sky coordinates using the input WCS (with Astropy's WCS class). The on-sky coordinates of these vertices are then mapped to output image coordinates using the newly constructed WCS. The footprint of the input pixel on the output frame is then represented as a Polygon object, using a Python package called Shapely. The coadd frame pixels within this footprint are also represented as Polygons, and the overlapping area between the input pixel and each output pixel is calculated. This step is computationally intensive, but provides a high level of flexibility and robustness to the coadd method, as the polygons are entirely flexible in shape and orientation. In particular, this allows us to implement a `Drizzle' factor, shrinking the size of the input pixels by a certain amount (typically to $70-80\%$ the original size) to increase the spatial sampling of the coadd \citep{Drizzle}. Time is also not a major constraint for the typical use-case of CWITools coadding, as it only needs to be performed a small number of times per target. That said, the process still only takes about twenty seconds to add 3-4 high-resolution KCWI cubes, including error propagation and masking. The contributions from the individual input cubes are weighted by exposure time, $E$. Thus, if we let the index $i$ iterate over the input cubes, and the indices $j, k$ iterate over the two spatial axes, such that $x_{ij}$ is the $j^{th}$ x-pixel of the $i^{th}$ cube, then the final coadded flux is given below in Equation~\ref{CWITools:Equation:CoaddFlux}.

\begin{equation}
    F_{coadd}(x, y) = \frac{ \sum_{i} E_i \big[ \sum_{j} \sum_{k} F_{in}(x_{ij}, y_{ik}) f(x, y, x_{ij}, y_{ik})\big]}{\sum_i E_i}
    \label{CWITools:Equation:CoaddFlux}
\end{equation}

Here, $f(x, y, x_{ij}, y_{ik})$ is the fraction of the footprint of the input pixel ($x_{ij}, y_{ik}$) that falls on the output pixel ($x,y$). Since the wavelength axes have been aligned, and the process here is applied at all wavelength layers, the third axis has simply been dropped from the notation. The propagated variance is then as shown in Equation~\ref{CWITools:Equation:CoaddVariance}.

\begin{equation}
    V_{coadd}(x, y) = \frac{ \sum_{i} E_i^2 \big[ \sum_{j} \sum_{k} V_{in}(x_{ij}, y_{ik}) f^2(x, y, x_{ij}, y_{ik})\big]}{(\sum_i E_i)^2}
    \label{CWITools:Equation:CoaddVariance}
\end{equation}

It should be noted that the interpolation involved in the wavelength alignment and flux redistribution introduces additional covariance in the coadded cube. While CWITools does not currently have a built-in way to calibrate covariance, a full section is dedicated to discussing this in \citet{OSullivan:2020a}. Some pixels at the edge of the coadded field of view may only be partially covered by the input data after all of the input cubes have been added. A threshold is (optionally) applied to reject any edge pixels with very low coverage, and referred to in the code as `pxthresh' - meaning pixel coverage threshold. Setting this to a high value (i.e. 0.9) will mean that only more-or-less fully covered pixels are included. \\ 

As a final step, empty rows, columns, and wavelength layers are trimmed from the coadded data. Here, a second (optional) threshold is applied based on exposure time. If the input data has any spatial dithering, some spaxels will have longer total exposure times than others. Spaxels with significantly lower exposure times will appear noisier and may complicate analysis. The parameter `expthresh' sets the minimum exposure time (as a fraction of the maximum) for a spaxel to be included in the data. This threshold is only applied as an extension of the trimming; that is, rows and columns on the edge of the field of view with lower exposure times will be trimmed from the data. It does not remove or mask regions throughout the data arbitrarily. As an example, if three equal-length exposures are taken with a 0, $-1''$, $+1''$ dithering pattern along the x-axis, the edge $-1''$ regions along that axis in the coadd will have fractional exposure times of $1/3$. Setting expthresh to 0.5 would eliminate these regions and keep only the overlapping central part of the field of view. Coadded cubes are saved by default using the same name as the ``.list'' file, though the output filename can be specified during usage.

\begin{figure*}[t]
    \centering
    \includegraphics[width=0.49\textwidth]{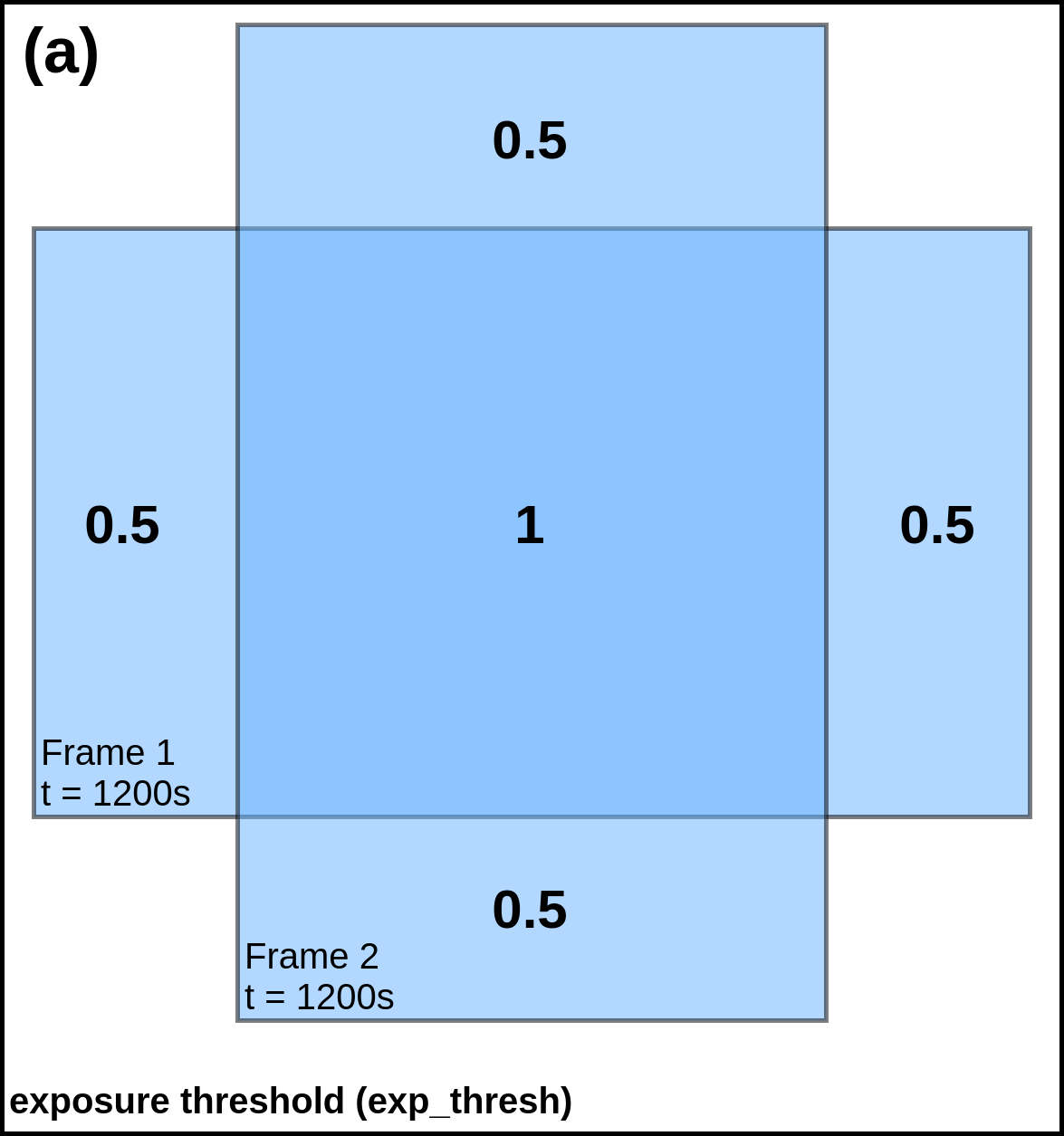}
    \includegraphics[width=0.49\textwidth]{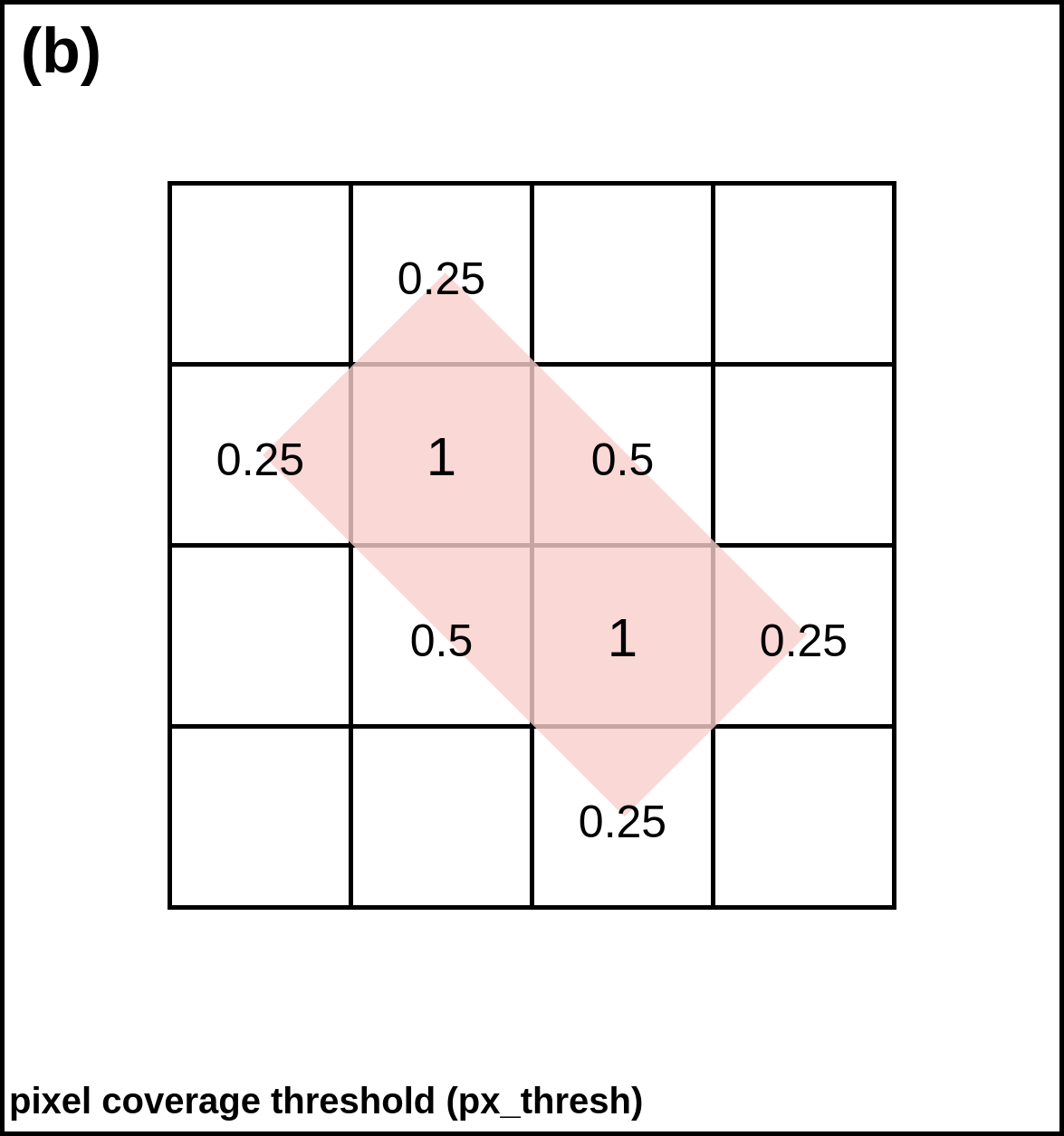}
    \caption{Illustration of the exposure and pixel coverage thresholds in CWITools' coadd function. Panel (a) shows an illustration of two overlapping fields of view with equal exposure time. The numbers in each area represent the local stacked exposure time relative to the maximum stacked exposure time. These are the values considered when applying the exposure threshold. Panel (b) illustrates the pixel coverage threshold. The white grid represents the coadd pixel grid, while the red rectangle represents the footprint of an input pixel.}
    \label{CWITools:Figure:Coadd}
\end{figure*}

\subsection{Variance Estimation and Scaling} \label{CWITools:Section:Variance}

While KDRP and PDRP produce 3D variance estimates for each exposure, there may occasionally be a need to estimate the variance from the data itself. For example, if some procedure is performed on the data for which the error propagation is prohibitively complex or if there is some problem with the data that affects the pipeline estimate, it may be preferable to estimate the variance empirically. In these cases, the basic approach, which is broadly similar to that used in \cite{Borisova:2016}, is to first estimate a 2D variance map by taking the variance along the z-axis and then scale that 2D variance map to match the noise properties of each wavelength layer in the data. This is complicated by the presence of real signal in the data so, as a first pass, let us assume that the cube is dominated by noise and that the number of voxels containing such signal is negligible. \\ 

The cube is first divided into bins of size $\Delta z$ pixels (i.e. wavelength layers). For each bin, the variance is taken along the z-axis to produce a local estimate of the (x,y) variance. Then, for each layer, the distribution of SNR values using this variance estimate is calculated. The following step relies on an assumption that the noise is Gaussian (or at least approximately Gaussian) in form as it involves the assertion that the distribution of SNR should follow a standard normal distribution ($\mu=0$, $\sigma=1$). This, in turn, relies on the assumption that (i) the distribution is dominated by background pixels and (ii) the noise within the background of the data is Gaussian in nature. These are both reasonable assumptions for a long (i.e. sky-limited) exposure containing no bright sources and with only a small fraction of the voxels/spaxels occupied by real emission. This assumption obviously breaks down under different circumstances, which we will discuss shortly. If the measured distribution of SNR values in this layer has standard deviation $\sigma_i$, then the variance rescaling factor is $r_{var}=1/\sigma_i^2$.   \\ 

\begin{figure*}[t]
    \centering
    \includegraphics[width=0.75\textwidth]{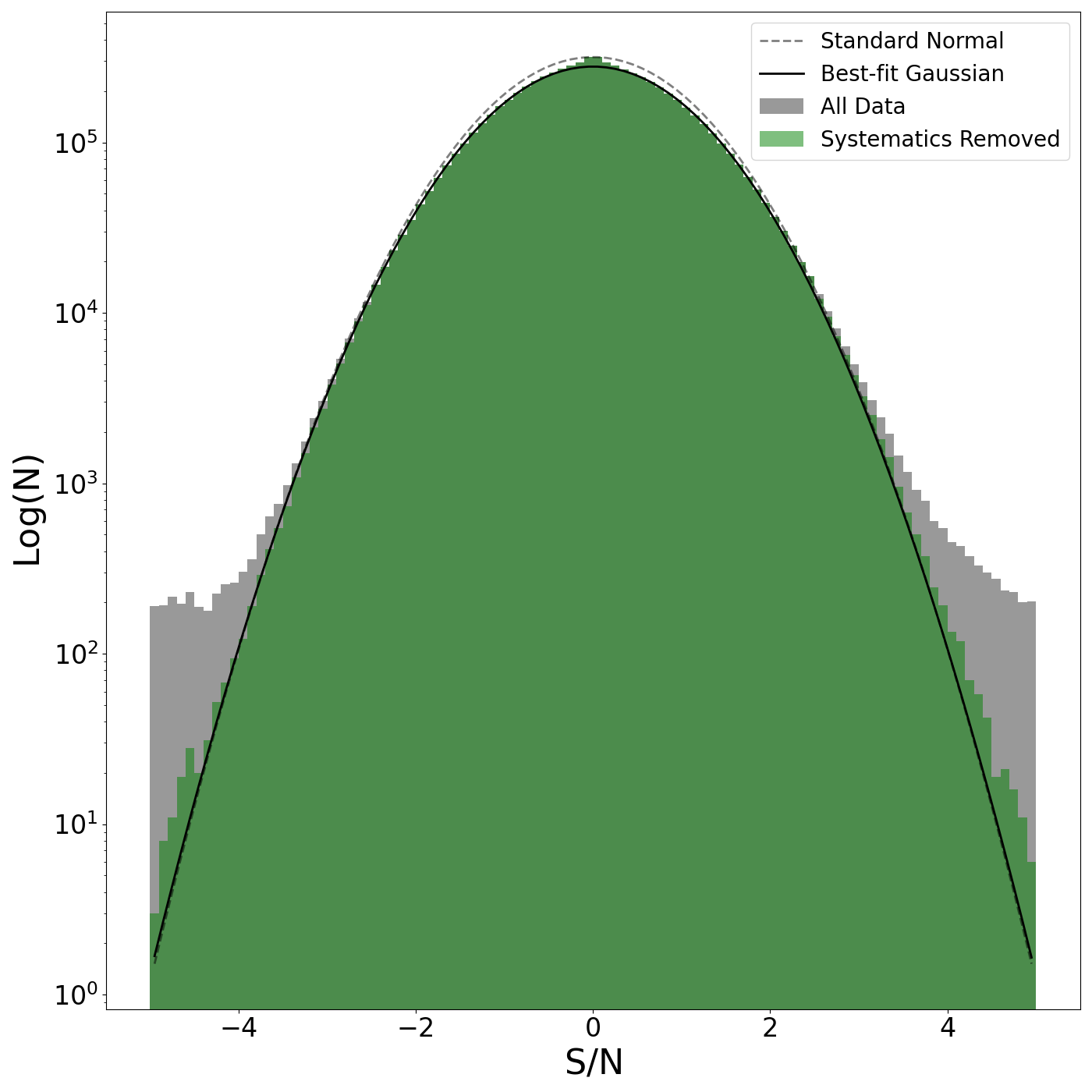}
    \caption{Variance scaling by assertion of a standard normal distribution in `background' regions. The grey shaded histogram shows the distribution of SNR values based on the input data and variance. Note the logarithmic scale on the y-axis. The green histogram shows the distribution of SNR values after large, contiguous 3D objects (either systematic residuals or real emission regions) have been detected and masked. The dashed black line shows a standard normal distribution, and the solid black line shows the best-fit Gaussian model used to calculate the re-scaling factor. }
    \label{CWITools:Figure:Variance}
\end{figure*}

The assumption of a standard-normal distribution only applies for background regions which are shot-noise limited. In the case where the input data contains large regions of bright emission, these regions must be masked and excluded from the SNR distribution which is used to calculate the scaling factor. Ultimately, there must still be a reasonably large background region - confidently free of real signal - to justify use of this method. Otherwise, this method should not be used to estimate the variance. If a mask is provided, two restrictions apply. The first is that within every z-axis bin, there must still be enough unmasked wavelength layers to obtain the variance along the z-axis in every spaxel. If some regions of the mask are very extended in wavelength, the $\Delta z$ parameter should be increased to ensure that enough unmasked layers remain in each bin. This, of course, reduces the accuracy of the local 2D variance estimate, but is a necessary step. The second restriction is that every wavelength layer must still contain a sufficient number of unmasked spaxels to obtain a reliable distribution of SNR values. If this is not the case, again, this method should not be used. As a last resort, if both of the above restrictions cannot be met, a single scaling factor can be applied to the entire variance cube by combining all background voxels into a single distribution.\\ 

If a user already has a variance cube, but believes it needs to be scaled (e.g. to account for covariance introduced by coadding), then the initial variance cube can be provided and only the rescaling part of the algorithm will be applied. Estimated and scaled variance cubes are saved by default with the extension ``.var.fits.''

\begin{figure*}[t]
    \centering
    \includegraphics[width=\textwidth]{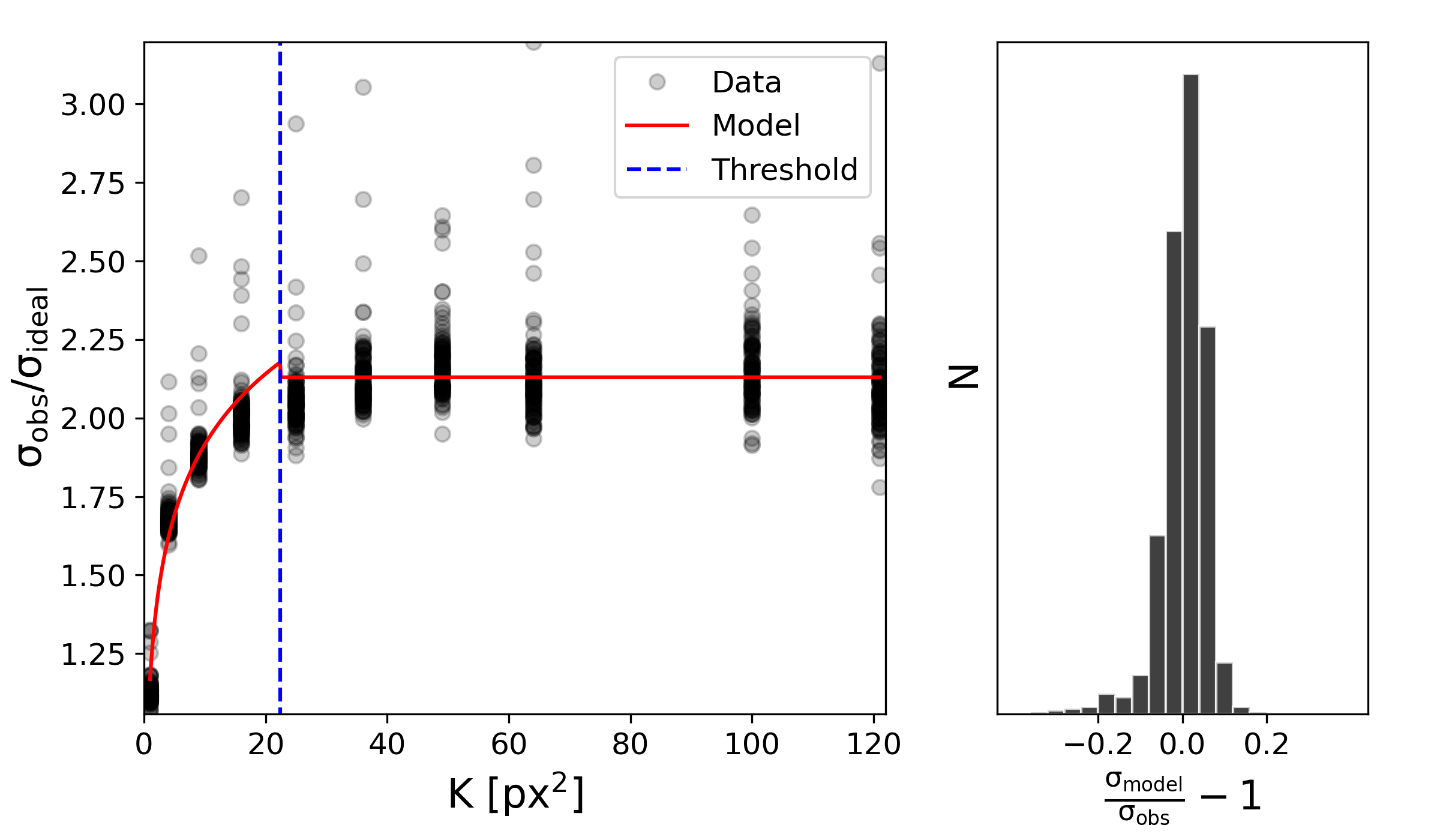}
    \caption{Calibration of covariance in a single KCWI data cube. The panel on the left shows the ratio between the observed noise and the propagated noise assuming no covariance ($\mathrm{\sigma_{obs}}$ and $\mathrm{\sigma_{ideal}}$) - after the binning the data - as a function of bin size. The red line shows the best-fit two-component model, with $\mathrm{\sigma_{obs}/\sigma_{ideal}= C (1 + \alpha \log(K))}$ for $\mathrm{K \leq K_{thresh}}$ and $\mathrm{\sigma_{obs}/\sigma_{ideal} = \beta}$ for $\mathrm{K > K_{thresh}}$. The right panel shows a histogram of the fractional residuals. }
    \label{CWITools:Figure:Covariance}
\end{figure*}

\subsection{Slice-to-Slice Scattered Light Correction}
In image-slicer integral field units, each slice of the field of view is sent along a different optical path. Part of the standard DRP's job is thus to correct for the differing relative response (i.e. throughput) of different slices, which can be caused by dust on the slices or pupil mirrors, or edge-of-field effects. One slice-to-slice correction that falls beyond the scope of the standard KCWI and PCWI DRPs involves scattered light. Slices containing very bright sources can sometimes contain an additional, relatively flat scattered light component across the slice. To remove this, CWITools runs through the 1D profile of each slice at each wavelength layer, estimates the background level, and subtracts it. The estimate is made by conservatively sigma-clipping the 1D profile to remove bright sources and then taking the median of the remaining pixels. This method is suitable for relatively clear fields with a single, very bright source. In fields with multiple sources, it can be difficult to obtain a reliable background estimate and this method should only be used with appropriate caution. Slice-corrected cubes are by default saved with the extension `sc.'

\subsection{Air-to-Vacuum and Heliocentric Corrections}
Lastly, the reduction module contains two common corrections for the wavelength axes of input data: conversion from air wavelengths to vacuum wavelengths, and a heliocentric velocity correction. For the former, CWITools uses an implementation from the package PyAstronomy to convert the wavelength axis. Because the correction from air to vacuum wavelengths depends on wavelength, the data must be interpolated onto the new wavelength grid, using either linear or cubic interpolation. Error propagation is not yet available for this function, so the variance should be rescaled or estimated anew after the application of this change. For the heliocentric correction, CWITools uses Astropy's SkyCoordinate class, and offers a choice between updating the header keywords to modify the wavelength axis, or to keep the original wavelength axis and shift the data using interpolation.

\section{Module: Extraction}

The extraction module could just as well be called the `isolation' module, as the ultimate goal is to isolate a specific signal, be it a point source, extended source, continuum emission, or line emission. While specifics may vary, as always, there are a few more or less ubiquitous steps in this process. First and foremost among them is the removal of point sources by modeling of the point-spread-function. Second is the removal of any unwanted component of the emission which is slowly varying both spatially and spectrally, referred to loosely as `background subtraction.' Masking, smoothing, and segmenting the data (into contiguous regions above a threshold) are also common steps towards isolating a signal. In this section, we describe the CWITools implementation of each of these.

\subsection{PSF Subtraction} 

PSF subtraction requires first modeling the PSF in 3D and then subtracting the model. Analytical PSF models such as Gaussian or Moffat profiles provide robustness of shape, which helps when trying to avoid overfitting, and are more well suited to fitting blended sources. However, the real instrument PSFs in PCWI and KCWI are more complex than a simple Gaussian or Moffat. As such, relying on these models for PSF subtraction leads to significant, systematic residuals. Systematic errors can be significantly worse than random error as they run the risk of creating false negatives and false positives. As such, the CWITools PSF-subtraction follows an empirical approach, some variant of which is widely used in existing observational work \citep{ArrigoniBattaia:2019a,Cai:2019,OSullivan:2020a}.\\ 

The most common reason for performing PSF subtraction in CWI data is to disentangle point sources and extended, nebular line emission. The key property of nebular line emission that enables this particular method is that it is spectrally confined to relatively narrow portions of the overall bandwidth. The empirical approach to building a 3D PSF model involves constructing a 2D model of the PSF by summing over wavelength layers which do not contain nebular emission (`continuum wavelengths'), then scaling it to match the PSF in each wavelength layer. The benefit of this approach is that arbitrarily complex instrument PSFs can be reliably subtracted, provided the shape does not change strongly as a function of wavelength. The most significant drawback of this approach is that it struggles to handle blended PSFs of two or more sources, as adjacent sources will be included in the empirical model. It is also not well suited to separating a diffuse/resolved continuum source from an unresolved continuum source. The way to achieve these goals following an empirical approach is to use an isolated source to obtain a PSF model. While this may be added as an option in a future update of CWITools, the current version focuses on extended nebular emission.\\

In the current implementation of this method, the user can specify a series of wavelength regions to exclude from WL images. For each wavelength layer, a new WL image is calculate by summing over a window of width $\delta\lambda_{WL}$ centered on the current layer. If the window is clipped on either side by the limits of the z-axis or there are masked layers within it, it is grown incrementally until the `effective' window size (i.e. the total \textit{useable} bandwidth) is equal to $\delta\lambda_{WL}$. This ensures that the number of wavelength layers summed for each WL image is consistent, ensuring that the SNR of the PSF model also remains roughly consistent (it will of course vary anyway depending on the spectrum of the continuum emission). Pixels within a radius of $r_{min}$ (default value is typically $r_{min}=1''$) from the center of the source are used to calculate a scaling factor for the PSF at each wavelength layer. The scaled model is then subtracted from the layer out to a radius of $r_{max}$, which is typically set to 2-3 times the seeing (i.e. $r_{max}\sim5''$). 

CWITools has two modes of PSF subtraction: 1D and 2D. In the 2D version, the above process takes place using full 2D white-light images and circular regions of radius $r_{min}$ and $r_{max}$. In the 1D method, the PSF models are created, scaled, and subtracted on a slice-by-slice basis. This is motivated by the fact that, for bright sources, there can be a significant scattered light component which is slice-dependent, and thus better fit by a model for that slice alone. The same $r_{min}$ and $r_{max}$ are used, only now in a 1D sense and for each slice. \\ 

\begin{figure*}[t]
    \centering
    \includegraphics[width=\textwidth]{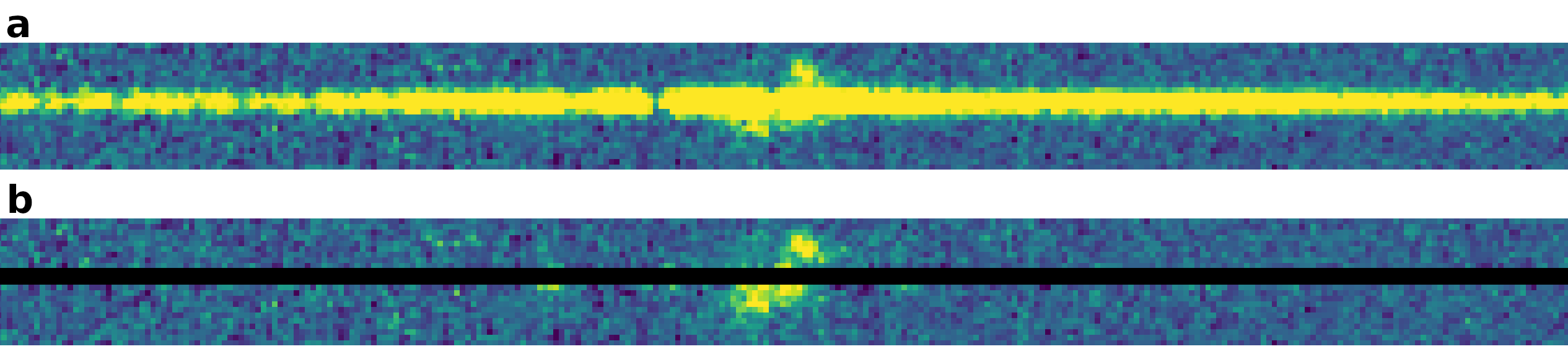}
    \caption{PSF-subtraction of a bright source to isolate extended emission. The top panel (a) shows a section of the 2D spectrum of a slice containing the bright source. The bottom panel shows the same 2D spectrum, with the same color map, after PSF subtraction. Bright, extended Ly$\alpha$ emission can be seen clearly after the subtraction. The small, bright spots to the left (blue) side of the extended emission are systematic residuals caused by sharp Ly$\alpha$ forest absorption features, where the PSF shape changes rapidly as a function of wavelength. The central pixels used to scale the PSF are masked, as these cannot be used to measure signal. }
    \label{CWITools:Figure:PSFSub}
\end{figure*}

In either method, variance data can be provided. If it is, the variance on the PSF model is calculated and error is propagated throughout the subtraction. Final output is saved with the extension ``.ps.fits'' (for PSF-Subtracted) and ``.ps.var.fits'' for the associated variance. Optionally, the PSF model can also be saved.

\subsection{Background Subtraction } 

As mentioned briefly above, the goal of background subtraction (BGSub) is to separate and remove any slowly varying component, spatially or spectrally. Examples of such signals include residuals left over after an imperfect sky subtraction or extended continuum emission from a (foreground) galaxy. As such, the term `background' is used quite loosely. There are many ways to approach removing background signals. CWITools has three methods to choose from at the time of writing: (i) polynomial spectral fitting, (ii) median filtering, and (iii) simple median subtraction. 

\subsubsection{BGSub Method 1: Polynomial Fitting}

\begin{figure*}[t]
    \centering
    \includegraphics[width=\textwidth]{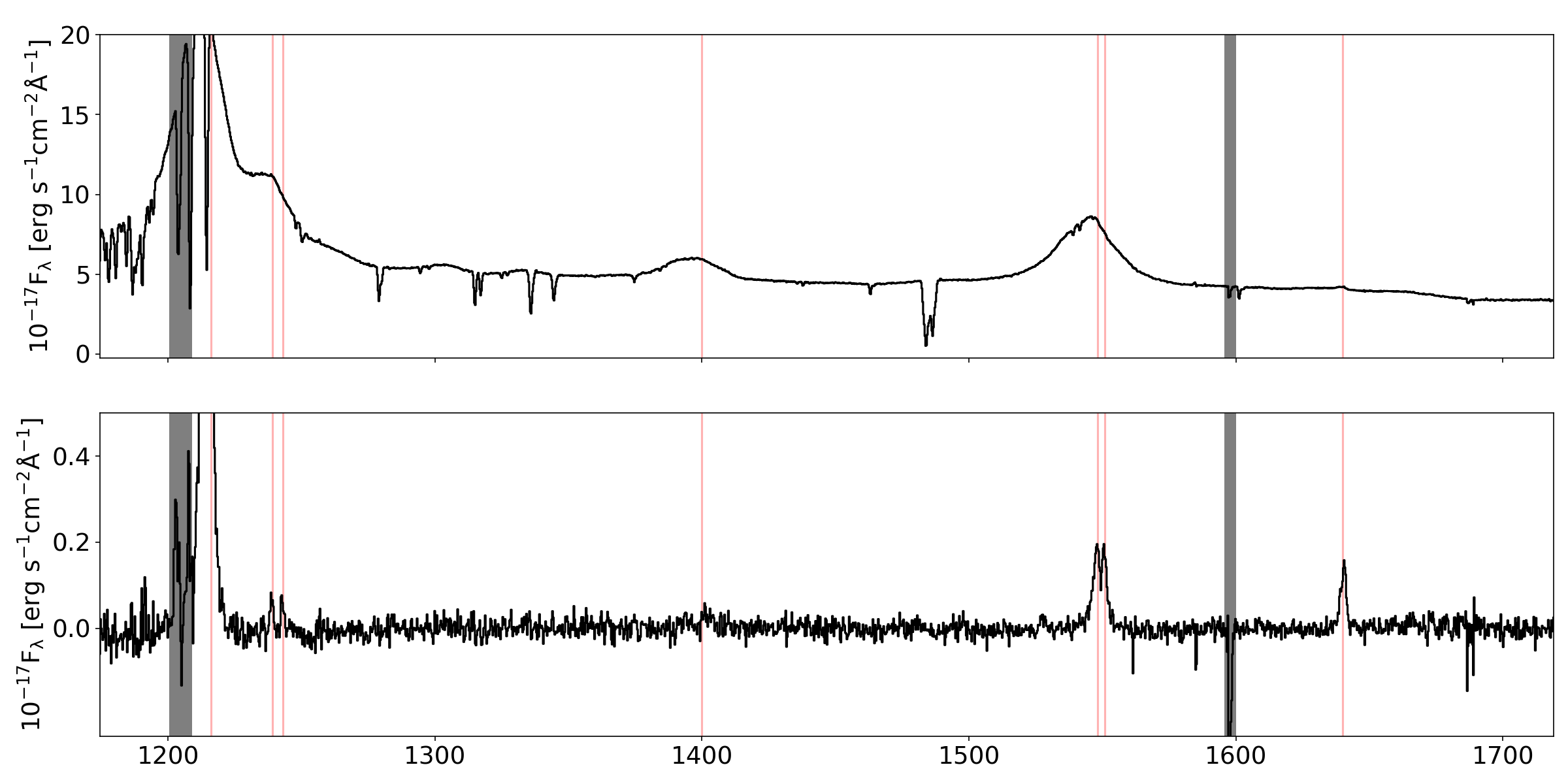}
    \caption{Integrated spectrum of a region before (top) and after (bottom) PSF and polynomial background subtraction. Black vertical bands indicate the locations of noisy residuals from extremely sharp absorption features or bright sky line, both of which can be masked. Vertical red lines, from left to right, indicate the positions of emission lines: HI Ly$\mathrm{\alpha}$ $\mathrm{\lambda}$1216, NV $\mathrm{\lambda\lambda}$1239, 1243, SiIV $\mathrm{\lambda}$1394, CIV $\mathrm{\lambda\lambda}$1548, 1551, HeII $\mathrm{\lambda}$1640. }
    \label{CWITools:Figure:SpecSub}
\end{figure*}

The `polyfit' method fits a low-order polynomial (i.e. polynomial degree $k_p\simeq1-5$) to the spectra in each spaxel. This method is probably the best choice for applications involving narrow-line nebular emission within data where the total bandwidth is large relative to the width of the emission. In such a scenario, the emission line features will be too small to be over-fit by such a slowly varying polynomial, and the fit will be dominated by continuum wavelengths. However, if the total bandwidth is similar to the width of the emission, even low-order polynomials will be more susceptible to over-fitting. In either case, wavelength regions known to contain emission lines can be masked by the user, ensuring that the polynomial is only fit to the background signal.  

If the background polynomial model is:
\begin{equation}
    p(k, \lambda) = \sum^{k}_{i=0} c_i \lambda^i,
\end{equation}
then, assuming that the wavelength of each layer is a well known quantity, and the only variance comes from uncertainty in the coefficients, the variance on the background model can be written as:

\begin{equation}
V(k, \lambda) = \sum_{q=0}^{q=k} \Big( \frac{\delta p(k, \lambda)}{\delta c_q} \cdot \delta c_q \Big)^2.
\end{equation}

The partial derivative expands to:
\begin{equation}
\frac{\delta p(k, \lambda)}{\delta c_q}  = \sum_{i=0}^{i=k} \Big( \lambda^i \frac{\delta c_i}{\delta c_q} + c_i \frac{\delta \lambda^i}{\delta c_q} \Big) = \sum_{i=0}^{i=k} \lambda^i \frac{\delta c_i}{\delta c_q},
\end{equation}

where again, wavelength is assumed to be a well known quantity. Plugging this in to the expression for the variance, we get:
\begin{equation}
V(k, \lambda) = \sum_{q=0}^{q=k} \Bigg( \sum_{i=0}^{i=k} \lambda^i \frac{\delta c_i}{\delta c_q} \cdot \delta c_q \Bigg)^2.
\end{equation}

The covariance matrix for the polynomial coefficients is returned by the polynomial fitting routine used (NumPy's polyfit). The off-diagonal covariance terms are typically very small compared to the diagonal terms. As such, the simplifying approximation can be made that the variables are independent, such that $\delta c_i/ \delta c_q = \delta^i_j$, where $\delta^i_j$ is the Kronecker delta function. This reduces the above expression to

\begin{equation}
V(k, \lambda) = \sum_{q=0}^{q=k} \lambda^{2q} (\delta c_q)^2 = \sum_{q=0}^{q=k} \lambda^{2q} \mathrm{Var}(c_q).
\end{equation}

The variance estimate can be re-scaled following Section~\ref{CWITools:Section:Variance} if needed to account for added covariance in the data.

\subsubsection{BGSub Method 2: Median Filtering}

Another common method to subtract background signals is median filtering along the wavelength/z-axis. The background is estimated by means of a median filter with window size $\Delta z \geq 5$, in pixels, wherein the value at each z-index is replaced with the local median. As in the polyfit method, wavelength regions can be masked to exclude them from contributing to the estimate of the local median. However, the window size must be sufficiently large that at least five unmasked pixels remain in the window, or a median is not well defined. While median filtering can be useful in scenarios where a polynomial fit performs poorly, median-filtered models are less well behaved in the sense that they can produce sharp, small-scale features and change discontinuously from one index to the next. Especially in the context of masking certain wavelengths, they should be used with caution and the background model (which can be saved as an option in CWITools) should be inspected. \\ 

The variance on the mean in a given window of size $N_z$ centered on index $i$, as a function of the existing variance estimates, is:

\begin{equation}
    V(N_z, i) = \frac{1}{N_z^2} \sum_{j=i-\Delta z /2}^{j=i+\Delta z/2} V_j
\end{equation}

where $V_j$ is the input variance at index j. The ratio of the variance on the mean to the variance on the median is $4n/(\pi(2n+1))$ where $N\equiv 2n + 1$ is the total sample size. This can be rewritten in terms of $N$ as $\pi N / (2(N-1))$. Thus, the variance on the \textit{median} background model is:

\begin{equation}
    V_{med}(N_z, i) = \frac{\pi}{2N_z(N_z-1)}\sum_{j=i-N_z/2}^{j=i+N_z/2} V_j.
\end{equation}

\subsubsection{BSub Method 3: Median Subtraction}

Where the median filtering method works along the z-axis, the simpler `medsub' method subtracts the spatial median at each wavelength layer. This is most useful in data that (i) is well flat-fielded and (ii) contains mostly background spaxels at all wavelength layers. The latter constraint can be relaxed if a mask is provided by the user to exclude non-background regions, but the remaining background region should still be sufficiently large at all wavelength layers to obtain a reliable median. Since the variance propagation on this is a simpler version of that described in the median filtering method above, it does not need to be described in detail again.

\subsection{Masking} 

Masking can be necessary after the subtraction of some point sources, where relatively small errors in the shape of the PSF near the core can still lead to loud residuals, or in order to mask the locations of foreground sources to exclude them from subsequent analysis. CWITools' extraction module contains a method for creating 2D binary masks using DS9 region files (extraction.get\_mask). The method takes in 2D image data to obtain the dimensions for the output mask. A user can also define a custom 1D (i.e. spectral), 2D (spatial), or 3D mask, and use the CWITools masking method to apply it to any compatible data. The mask and data can both be 3D, both 2D, 3D data with a 2D mask (in which case the mask is treated as a spatial mask and applied at all wavelength layers), or 3D data with a 1D mask (in which case the mask is treated as a spectral mask). Masked data is saved by default with a ``.M.fits'' file extension. 

\subsection{Smoothing}

While there are many existing smoothing and filtering methods available from libraries such as NumPy and SciPy, three custom methods are included in CWITools to allow for error propagation (by squaring the smoothing kernel). The main smoothing method is a generic one which applies smoothing along a requested subset of axes for 1D, 2D, or 3D input data. For 2D or 3D data, smoothing can be applied along any subset of the axes, and is implemented as a series of convolutions with 1D smoothing kernels. This is permitted by the fact that the two available kernels are Gaussian and Box kernels, both of which are separable into 1D components. Smoothing along the z-axis with a kernel of size $N_z$ is described mathematically as:

\begin{equation}
    F'(x, y, z) = \frac{\sum_{i=-N_z/2}^{i=+N_z/2} F(x, y, z+i)K_{1D}(i)}{\sum_i K_{1D}(i)}.
\end{equation}

The naive error propagation (i.e. ignoring the effect of covariance) follows:

\begin{equation}
    V'(x, y, z) = \frac{\sum_{i=-N_z/2}^{i=+N_z/2} V(x, y, z+i)K^2_{1D}(i)}{(\sum_i K_{1D}(i))^2}.
\end{equation}

As mentioned before, \citet{OSullivan:2020a} contains a full discussion of bootstrapping variance estimates to include covariance in PCWI data, following the lead of earlier work by the Calar Alto Legacy Integral Field Area survey \citep{Husemann:2013} and the SDSS-IV MaNGA IFU Galaxy Survey \citep{Law:2016}. CWITools does not yet have a built-in implementation of this method, but it is planned for a future release. \\ 

Two other methods are included as faster, stripped-down alternatives for the most common smoothing applications: spatial smoothing and wavelength smoothing. These make use of a faster convolution method ('SciPy.ndimage.convolve') and have simpler usage in that they always take 3D data and do not require the user to specify the axes.

\subsection{Segmentation}

While there are any number of additional steps that may be required depending on the application, a common final step in the extraction process is segmentation: dividing the cube into contiguous 3D regions above a certain threshold. The CWITools implementation of this is built using an existing  segmentation routine from the popular package `Scikit Image' (specifically skimage.measure.label). The function allows the user to set the segmentation threshold either in absolute physical units or in terms of SNR. It also provides the ability to limit the segmentation to specific wavelength ranges of interest and set a lower bound on the number of voxels required for a region to be included. Wavelength ranges containing common nebular emission lines can be included by providing the redshift of suspected emission and a velocity window to explore around each line. Similarly, known bright sky lines in the Palomar/Keck sky spectra can be excluded automatically. Custom ranges to include or exclude can also be provided by the user (e.g. to rule out bad wavelength regions or extend the velocity range of an extra broad line). The output of this method - a cube containing labelled regions which I call the `object' cube as a shorthand - can be used in synthesizing results and making measurements for specific objects later on. 

\section{Module: Synthesis}\label{CWITools:Section:Synthesis}
The synthesis module takes 3D data products as input and returns vectors or arrays containing scientifically relevant results such as radial profiles or velocity maps. `Object' cubes produced by segmentation can be used for many of these functions to generate such results from specific regions only. 

\subsection{White-Light Images} 
Generation of white-light (WL) images is a straight forward process: a 2D image is formed by summing over a broad wavelength range. Two sets of wavelength regions should generally be excluded in doing so: bright sky lines and any nebular emission lines present in the data. The user can specify wavelength regions to mask, and also select an option to automatically mask some known sky lines which are built into the package (stored in the `data/sky/' subdirectory in the installation directory). The wavelength region used should also be restricted to the `WAVGOOD' region indicated by the header information, but usually the data will have been cropped to this range before generating WL images. The input is generally assumed to be in the standard KCWI/PCWI units of `FLAM16' - meaning $10^{16}\times \mathrm{erg~ s^{-1} cm^{-2} \text{\AA}^{-1}}$. If this is the input unit (i.e. if the keyword `BUNIT' is set to `FLAM16'), the WL image is converted to surface-brightness units ($10^{16}\times \mathrm{erg~ s^{-1} cm^{-2} arcsec^{-2}}$) following\\ 

\begin{equation}
    \mathrm{SB}_{16}(x,y) = \frac{\Delta \lambda}{(\Delta \theta)^2} \sum_{z} F_{\lambda, 16}(x,y,z),
\end{equation}

where $(\delta \theta)^2$ is the size of the input spaxels in units of square arcseconds, and $\Delta \lambda$ is the size of each wavelength layer in units of Angstrom. Error is propagated by summing the variance data over the same wavelength layers and squaring the multiplicative term. 

\begin{equation}
    \mathrm{Var(SB}_{16})(x,y) = \Bigg(\frac{\Delta \lambda}{(\Delta \theta)^2}\Bigg)^2 \sum_{z} V_{\lambda, 16}(x,y,z).
\end{equation}

\subsection{Pseudo Narrowband Images}

A pseudo-Narrowband (pNB) image is a Narrowband image formed by summing wavelength layers of a datacube. The synthesis module has a method for generating pNB images with or without the subtraction of a local white-light image. The user provides the central wavelength and bandwidth of the desired pNB image, and is returned the image, an estimate of variance on the image, a local white-light image, and the variance on the white-light image. The variance estimates are derived from 3D variance cubes, if provided, or by taking the variance along the z-axis in the input data, if not. Optionally, the user can request white-light subtraction, in which case the white-light image is scaled and subtracted from the pNB image. The scaling factor is calculated using a user-provided location, typically the location of a bright, central continuum source to be subtracted. \\ 

This tool is useful as an initial exploration of a data cube. By generating a series of these images at different wavelengths, one can form a channel map. Channel maps are useful tools in exploring IFU data, especially when the nature (i.e. center and width) of any emission is not already well known, and the signal may be too faint to detect easily on a voxel-by-voxel basis (see \cite{OSullivan:2020a}). In the case of non-detections, this tool provides a useful way to obtain upper limits on the surface brightness of suspected emission.

A commonly studied property of both galaxies and extended emission is the circularly average radial surface brightness profile, usually as a function of some physical distance (i.e. comoving or proper kiloparsecs). CWITools provides a convenient function for the calculation of a radial surface brightness profile, where the radius can be returned in units of pixels, arcseconds, comoving kiloparsecs, or proper kiloparsecs. The radial bins are defined by four parameters: a minimum radius, a maximum radius, the number of (equally spaced) bins, and a scale setting which determines whether to make the bins of equal size in linear space or log space. The user provides a surface brightness map and central location as input. The surface brightness profile can be easily obtained using either the synthesis.pseudo\_nb method or synthesis.obj\_sb method (below) which measures the surface brightness of a defined 3D region. 

\subsection{Object Surface Brightness, Spectra and Moments}
In many IFU studies, though especially those regarding extended nebular emission, 3D object masks (generated by segmentation - see extraction.segment) are a central feature. These masks contain integer-labelled, contiguous 3D regions which we refer to here as `objects.' Once one has a 3D region delineating the object of their study, it becomes trivial to generate useful products. CWITools has a number of functions which accept 3D object masks, along with specified object IDs (i.e. the number of the region to study). Specifically, the synthesis module contains a method for generating: (i) a surface brightness map of an object (synthesis.obj\_sb), (ii) an integrated 1D spectrum of an object (synthesis.obj\_spec), and (iii) maps of the first two z-moments (i.e. velocity and dispersion) for an object (synthesis.obj\_moments). \\ 

Object surface brightness maps are obtained by setting all non-object voxels to zero, summing the cube along the z-axis, and applying a conversion from units of $\mathrm{F_\lambda}$ (i.e. $\mathrm{erg/s/cm2/\text{\AA}}$) to units of surface brightness ($\mathrm{erg/s/cm2/arcsec2}$). One-dimensional spectra are similarly obtained by setting all non-object voxels to zero and summing along the two spatial axes. In this case, no unit conversion is required. The user can decide whether to apply the 3D mask in full or to sum the full spectra of all spaxels within the object boundary. The latter method can be useful for the purpose of presentation in that it shows the background noise throughout the rest of the spectrum, and in recovering the full shape of emission lines, as the thresholding step used to obtain 3D masks necessarily cuts out the faint wings of a profile once they fall below the noise level. Finally, 2D maps of the first and second moments in wavelength are obtained through a straight-forward moments calculation (see measurement.first\_moment and measurement.second\_moment for details and uncertainty propagation). In the case where moments are being calculated using a 3D mask, it is assumed that all object voxels contain positive flux, so no further threshold is applied and the `closing window' method is not used.

\section{Module: Modeling}
The modeling module provides wrapper functions for some common models, model fitting methods, model comparison, and some other miscellaneous useful tools. While there is a wealth of existing modeling functionalities available from Astropy and other packages, the CWITools modeling module contains wrappers for models commonly used in PCWI/KCWI data analysis with a self-consistent syntax for use in model fitting and evaluation. At the moment, all of the models and modeling functions are one dimensional, as the main applications considered are the fitting of emission lines and surface brightness profiles. Future updates, beyond the initial release, may include additional models, such as 2D kinematic or surface brightness models. 

\subsection{Wrappers for Models and Fitting}
Model fitting within CWITools is done by minimizing a residual sum of squares (RSS) function using SciPy's implementation of differential evolution as the default optimizer. Differential evolution is a stochastic method of optimization, which is less susceptible to local minima than standard gradient descent methods. This can be of importance in fitting models in the presence of significant noise, where the cost function is not smooth. 

Differential Evolution (in scipy.optimize), like many other available optimization methods, finds the minimum of an objective function of the form $f(p, [\mathrm{args}])$, where $p$ is the vector of free parameters to be optimized and $\mathrm{[args]}$ is a sequence of any additional, fixed parameters required for the function. Since CWITools uses a least-squares approach, the objective function is one which computes the residual sum-of-squares (RSS) for any given model, set of model parameters, and input data. This flexible RSS method (modeling.rss\_func) has the form $\mathrm{rss\_func(}p, f, x, y)$, where $p$ is the vector of model parameters, $f$ is the model function, and $x$ and $y$ are the input data. The model function itself must be of the form $f(p, x)$. For convenience, CWITools has a number of common functions written in this form, with more to be added later. The current library of functions includes a $1D$ Voigt, Gaussian, and Moffat profiles for line-profile or PSF fitting. For the fitting of radial profiles, 1D Sersic, Exponential, and Power-law profiles are included. 

\subsection{Model Comparison}
Model comparison can be performed using one of two information criteria: the Akaike Information Criterion (AIC) and the Bayesian Information Criterion (BIC). Both the AIC and BIC indicate the \textit{relative} likelihood that a given model is the best representation of the observed data out of all models considered, taking into account the degrees of freedom of the models. The AIC is calculated as
\begin{equation}
\mathrm{AIC} = 2k + n \ln(\mathrm{RSS}),
\end{equation}
where $k$ is the number of parameters, $n$ is the number of data points used in fitting, and $\mathrm{RSS}$ is the residual sum of squares. A lower score is better for both the AIC and BIC. The AIC also has a correction for small sample sizes, denoted as AICc:
\begin{equation}
\mathrm{AICc} = \mathrm{AIC} + \frac{2k^2 + 2k}{n - k - 1}.
\end{equation}
The correction term vanishes as $n$ approaches infinity. This term is always included in the CWITools implementation of the AIC. The BIC is calculated as:
\begin{equation}
\mathrm{BIC} =  k\ln(n) +  n\ln\Big(\frac{\mathrm{RSS}}{n}\Big).
\end{equation}
A set of AIC or BIC values can be converted into a set of weights indicating the relative likelihood of each model. Following \cite{Wagenmaker:2004}, these weights can be calculated as:
\begin{equation}
    w_i = \frac{\exp{(-\frac{1}{2}\Delta_i(\mathrm{BIC}))}}{\Sigma_j \exp{(-\frac{1}{2}\Delta_j(\mathrm{BIC}))}}.
\end{equation}

It is important to note that these weights indicate \textit{relative} likelihoods, with respect only to the other models considered. The scientific significance of such relative likelihoods therefore depends strongly on the total set of models considered. As a random example, it would be misleading to claim an absorption line is Gaussian in shape if the only models considered were a Gaussian model and a flat continuum model, but if other common line shapes (e.g. Lorentzian, Voigt) were considered and a Gaussian still had the lowest AIC/BIC, then it may be a reasonable claim. 

\begin{figure*}[t]
    \centering
    \includegraphics[width=\textwidth]{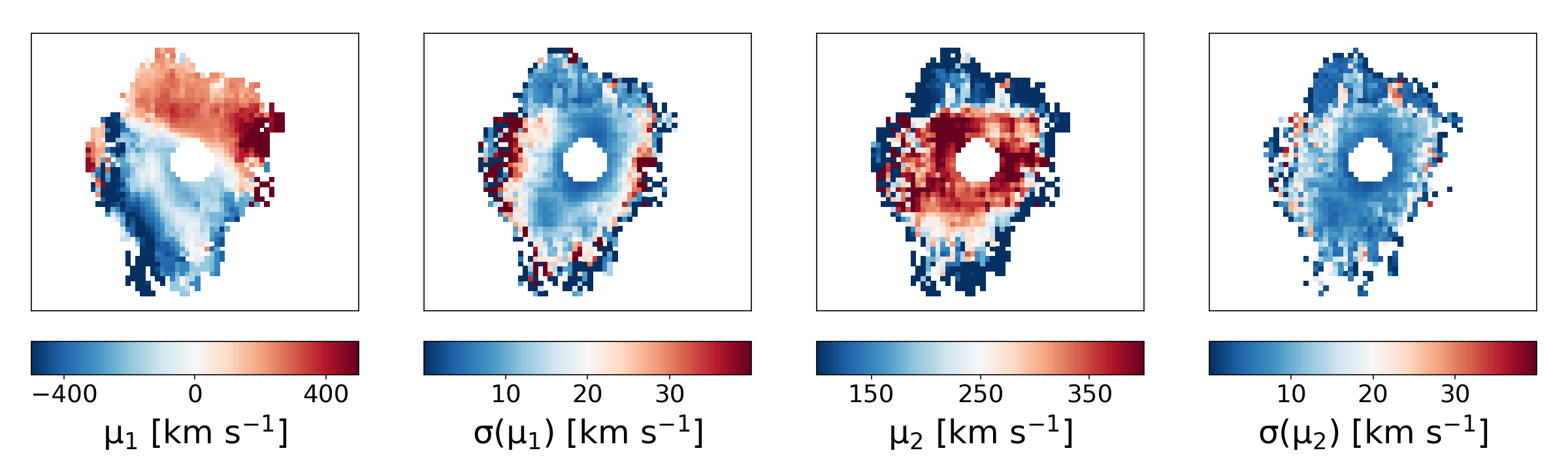}
    \caption{z-Moment maps, shown in units of $\mathrm{km~^{-1}}$, calculated using CWITools. From left to right: first moment, propagated error on the first moment, second moment, and propagated error on the second moment.}
    \label{CWITools:Figure:Moments}
\end{figure*}

\section{Module: Measurement}

\subsection{First and Second Moments}
In context of IFU data analysis, moments are typically calculated along the z-axis to derive velocity and dispersion. For that reason, we focus here on moments calculated along the z axis in wavelength units. The measurement module contains implementations of these moments calculations that accept a 1D wavelength axis and spectrum as input. The synthesis module contains a method which builds on this to create two-dimensional moment maps for 3D objects. The first moment of a 1D spectrum with wavelength $\lambda$ and flux $F$ is calculated as follows:
\begin{equation}
    \mu_1 = \frac{\sum_i \lambda_k F_k}{\sum_k F_k}.
\end{equation}\label{CWITools:Equation:Mu1}
As a short-hand, let us refer to the numerator here as $N_1$ and the denominator as $D$. The error on this calculation can then be shown to be
\begin{equation}
    \sigma(\mu_1) = \sqrt{\sum_i \frac{(\lambda_i D - N_1)^2}{D^4} V_i},
\end{equation}
where $V_i=\sigma^2(F_i)$ is the variance on the flux. The second moment is calculated as follows:
\begin{equation}
    \mu_2 =  \sqrt{\frac{\sum_i (\lambda_k -\mu_1)^2 F_k}{\sum_k F_k}}.
\end{equation}
Let us refer to the numerator this time as $N_2$. The denominator is the same as in Equation~\ref{CWITools:Equation:Mu1}. The error on this calculation can be shown to be:
\begin{equation}
    \sigma(\mu_2) = \frac{1}{2\mu_2} \sqrt{\sum_i \frac{((\lambda_i - \mu_1)^2 D - N_2)^2}{D^4} V_i}.
\end{equation}
It should be noted that the numerators here will have slightly higher variance due to the covariance between adjacent pixels. This can be taken into account following Section~\ref{CWITools:Section:Variance}.  

\subsection{Integrated Luminosity} 
The integrated luminosity can be calculated for 1D, 2D, or 3D data and an optional object mask of the same dimensions. If no object mask is provided, all of the input data is summed. If the input is two-dimensional, the input units are assumed to be those of surface brightness and the total flux, $F_{tot}$ is calculated following:
\begin{equation}
   \mathrm{ F_{tot} = (\Delta\theta)^2  \Sigma_{x}\Sigma_y SB(x,y)M(x,y) }
\end{equation}
where $\Delta\theta$ is the angular area in units of $\mathrm{arcsec^2}$ and $\mathrm{M(x,y)}$ is the 2D binary mask. If the input is 3D, the input unit is assumed to be units of $\mathrm{F_\lambda}$, and the luminosity is calculated as:
\begin{equation}
   \mathrm{ F_{tot} = (\Delta \lambda) \sum_{x}\sum_y\sum_z F_\lambda(x,y,z)M(x,y,z)},
\end{equation}
where $\Delta\lambda$ is the size of the wavelength layers, usually in Angstrom. Finally, if the input is one dimensional, the luminosity is calculated as:
\begin{equation}
   \mathrm{ F_{tot} = (\Delta \lambda) \sum_z F_\lambda(z)M(z)}.
\end{equation}
The total flux is then converted to luminosity using the luminosity distance, $\mathrm{D_L(z)}$:
\begin{equation}
   \mathrm{ L_{tot} = 4\pi D_L^2(z) F_{tot}}
\end{equation}
The error on the luminosity is obtained by summing the variance and squaring the multiplicative terms. The usual caveat regarding covariance (Section ~\ref{CWITools:Section:Variance}) applies here, and should be considered before using the estimated error.

\begin{figure*}[t]
    \centering
    \includegraphics[width=0.38\textwidth]{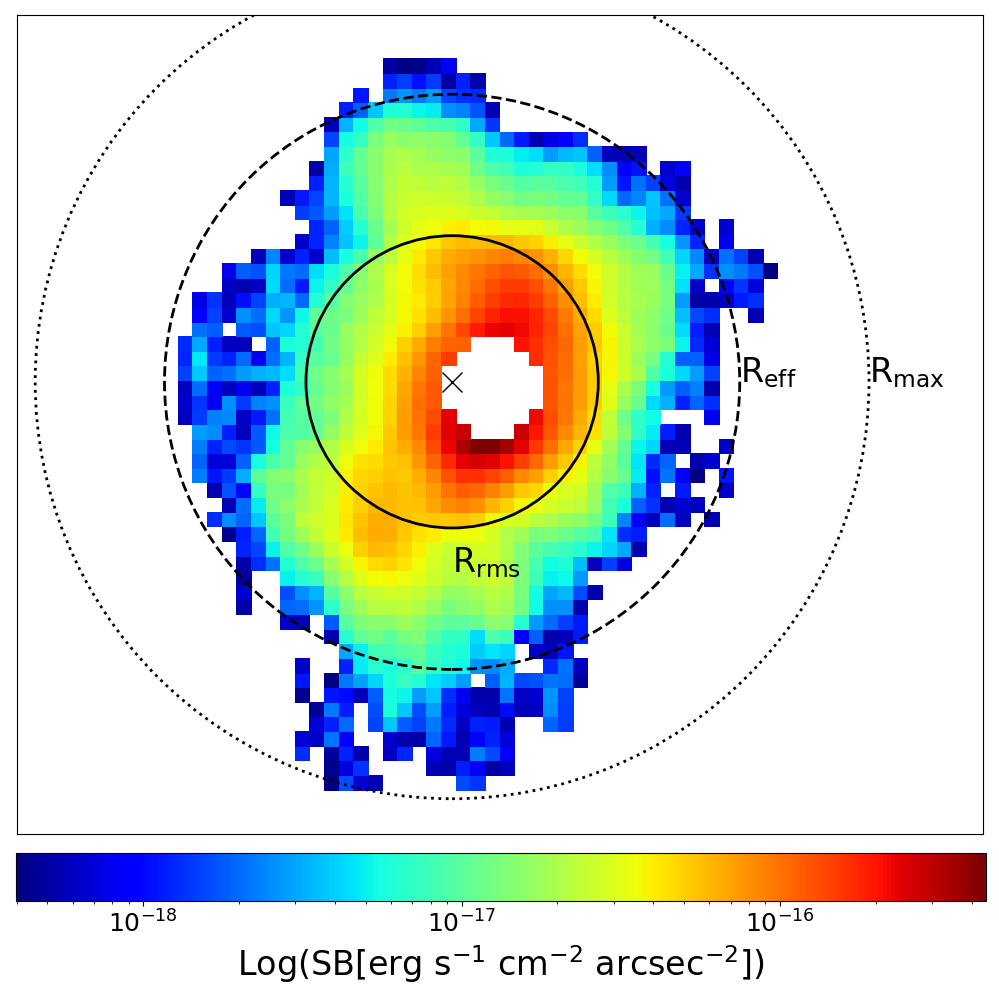}
    \includegraphics[width=0.61\textwidth]{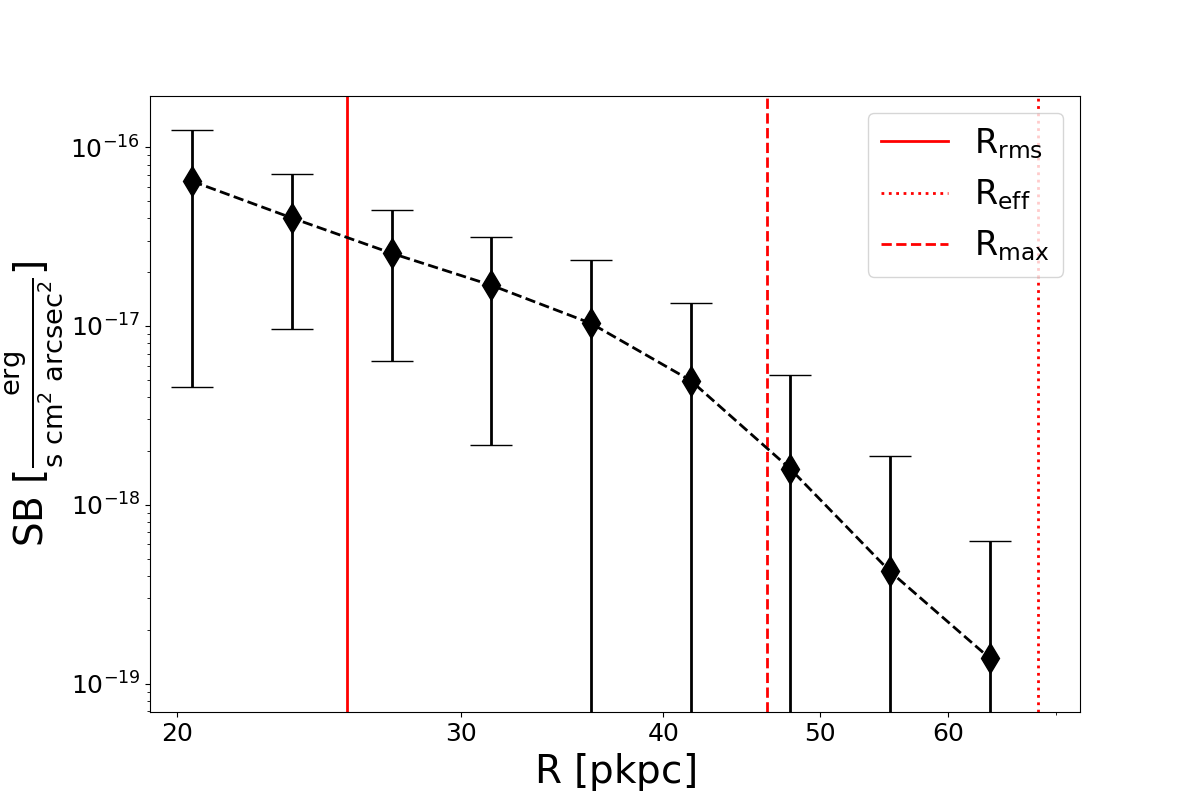}
    \caption{Left: an illustration of the three characteristic radii provided by the measurements module. From largest to smallest, they are: (i) the maximum radius ($\mathrm{R_{max}}$ - dotted circle), (ii) the effective radius ($\mathrm{R_{eff}}$ - dashed circle), and (iii) the flux-weighted RMS radius ($\mathrm{R_{rms}}$ - solid circle). Right: a radial profile generated by the synthesis module for the same nebula. The same three radii are shown on this axis.}
    \label{CWITools:Figure:Radii}
\end{figure*}

\subsection{Characteristic Sizes}

CWITools provides a number of ways to measure the size of a 2D or 3D object. First and foremost among them is area, either units of square pixels or square arcseconds. Several characteristic radii are also defined, as in \cite{OSullivan:2020a}; effective radius ($\mathrm{R_{eff}}$), maximum radius ($\mathrm{R_{max}}$), and RMS radius ($\mathrm{R_{rms}}$), defined as follows:
\begin{equation}
    \mathrm{R_{eff}} = \sqrt{A/\pi}
\end{equation}
\begin{equation}
    \mathrm{R_{max}} = \max [|r_{ij} - r_c|\times M_{ij}]
\end{equation}
\begin{equation}
    \mathrm{R_{rms}} = \sqrt{\frac{\sum_i \sum_j |r_{ij} - r_c|^2 F_{ij}M_{ij}}{\sum_i \sum_j F_{ij}M_{ij}}}
\end{equation}
where the indices $i$ and $j$ iterate over the spatial axes, $r_{ij}$ is the 2D vector from the image origin to the point $(x_i, y_j)$, $r_c$ is the vector from the origin to the flux-weighted centroid of the nebula, $F_{ij}$ is the total flux or surface-brightness at position $(x_i, y_j)$, and $M_{ij}$ is the 2D binary object mask, such that $M_{ij} = 1$ for object spaxels. If 3D data and object masks are provided, $F_{ij}$ and $M_{ij}$ are formed by taking the sum and maximum value along the z-axis, respectively. Each measurement, taken alone, serves as a useful reference for a different aspect of the object size, but lacks any information about shape. $\mathrm{R_{eff}}$ serves as a convenient proxy for total measured size, $\mathrm{R_{max}}$ describes the maximum extent of the nebula in any direction from its flux-weighted centroid, and $\mathrm{R_{rms}}$ provides a characteristic scale at which the emission is concentrated. 

\subsection{Asymmetry and Eccentricity} 
Another commonly used measurement in describing the morphology of an object is its asymmetry or eccentricity. The asymmetry parameter, $\alpha$ reflects the minor axis to major axis ratio of the emission. The calculation of this parameter, used in \cite{OSullivan:2020a}, is taken from \citet{ArrigoniBattaia:2016}, who in turn based it on work by \cite{Stoughton:2002}. It derives the parameter $\alpha$ from second order spatial moments defined as:
\begin{equation}
    M_{xx} = \Big\langle \frac{(x - x_{\mathrm{c}})^2}{r^2} \Big\rangle_f,
\end{equation}

\begin{equation}
    M_{yy} = \Big\langle \frac{(y - y_{\mathrm{c}})^2}{r^2} \Big\rangle_f,
\end{equation}
and
\begin{equation}
    M_{xy} = \Big\langle \frac{(y - y_{\mathrm{c}})(x - x_{\mathrm{c}})}{r^2} \Big\rangle_f,
\end{equation}

where $(x_c, y_c)$ is the flux-weighted centroid of the emission. As an aside, CWITools' measurement module contains a wrapper to enable arbitrary moment calculations of the form $M(xx, yy, p, q, f) = \langle xx^p yy^q \rangle_f$, where $xx$ and $yy$ are 2D mesh-grids of x-position and y-position, and $f$ is a 2D grid of flux-like weights. These second order moments are used to define the terms
\begin{equation}
    Q \equiv M_{xx} - M_{yy}
\end{equation}
\begin{equation}
    U \equiv 2M_{xy} 
\end{equation}
which are then used to derive the asymmetry
\begin{equation}
\alpha  = b/a = \frac{1 - \sqrt{Q^2 + U^2}}{1 + \sqrt{Q^2 + U^2}}.
\end{equation}
The `elliptical' eccentricity, which is another representation of the same thing, is defined as 
\begin{equation}
    e \equiv \sqrt{1 - \alpha^2}.
\end{equation}
Both of the above functions accept 2D or 3D data as input. If an object mask is not provided, all input data is used in the above calculation. This is generally only recommended if the input data already contains only the isolated 2D or 3D signal (i.e. an object mask has been applied by some means).

\subsection{Specific Angular Momentum}
One commonly studied phenomenon, in terms of kinematics, is the presence of structured kinematic shear. Shear-like features can arise from a number of different physical phenomena including inflows, outflows, and galactic rotation. Specific angular momentum is typically defined as angular momentum per unit mass, and has a dimensionality of square distance per unit time (e.g. $\mathrm{m^2 s^{-1}}$). Here, we use the flux of an object as a proxy for mass, and calculate the weighted average over the entire object. We also bear in mind that, in IFU data, we do not have full 3D vectors for the distance; rather, we have only projected radius and line-of-sight velocity. Thus to be more precise in our wording, the measurement is of the flux-weighted average \textit{projected specific} angular momentum of an object. This is defined as:
\begin{equation}
    \langle~\vec{j}~\rangle_f = \frac{\sum_x \sum_y F(x,y)\vec{R}_{\perp}(x,y)\times \vec{v}_{z}(x,y)}{\sum_x \sum_y F(x,y)}
\end{equation}
where $\vec{R}_{\perp}(x,y)$ is the projected radius, in $\mathrm{pkpc}$, from the flux-weighted centroid of the nebula to the point $(x,y)$, $F(x,y)$ is the flux at that point, and $\vec{v}_z(x,y)$ is the line-of-sight velocity, in $\mathrm{km~s^{-1}}$. The units of the measurement are thus $\mathrm{pkpc km s^{-1}}$. This measurement provides a useful, quantitative insight into whether the kinematics of an object appear to be dominated by a structured velocity gradient (which aligns with the distribution of flux to produce a large value) or whether the kinematics appear to be primarily noisy or chaotic.

\section{Summary and Discussion}
We have presented here a configurable pipeline for the extraction, modeling and measurement of signals in three-dimensional integral field spectroscopy data, and shown its application on extracting extended Ly$\alpha$ emission from a high-redshift QSO. CWITools provides a comprehensive and flexible suite of tools for correcting, coadding, and analyzing KCWI and PCWI data cubes. Here, we discuss briefly the extension of CWITools to other instruments and general future work.

\subsection{Extension to Other IFUs}
CWITools, as discussed, is built specifically with the Cosmic Web Imager instruments at Keck and Palomar observatory in mind. That being said, there is much that IFS data has in common, regardless of which instrument produces it. Certain header keywords are FITS standards, and others are not. Certain methods (e.g. polynomial subtraction) may apply universally, while others (e.g. analytical PSF modeling) may depend on instrument specifics. It is thus worth a quick look at the areas in which CWITools becomes instrument-specific, and how it could be adapted to other instruments.

The majority of methods in CWITools can be applied to any input containing the same data structure (a wavelength axis and two spatial axes). However, there are a few areas in which the specifics of the CWI format feature strongly. First and foremost among these is the nature of the file-types saved by the standard DRP. In KDRP and PDRP, for a single exposure, a separate data cube is saved for intensity, background, variance, and masks. As a counter example, the data cubes produced by MUSE store the variance associated with an intensity cube as a second HDU within the same file. This represents the biggest challenge in using CWITools for other instruments. However, in lieu of a package update with flexible input in every method, it can be patched by writing a tool to convert data from one instrument to the format of another. For example, a MUSE data cube could be loaded, written to separate intensity and variance files, and then used as input to CWITools. As such, this problem is easy to solve with a single function that converts data formats. This method can also be applied to data structures and header keywords; a specific function can be written to re-order axes and re-name keywords. Once this function is written, it is trivial to apply it to data and then continue using CWITools as on CWI data.

The next category of incompatibilities arises from package data that is specific to one instrument. For example, CWITools stores a list of known bright sky-lines at Palomar and Keck observatories. If certain tools were to be applied to MUSE data, a list of bright sky lines at European Southern Observatory would need to be added. CWITools also contains information about the different gratings and slicers available to PCWI and KCWI. These are used to estimate the spatial and spectral resolution. Again, this information would need to be added to make the same function available to another instrument. However, this amounts to simple data entry and is not a major obstacle. 

On the surface, it seems almost trivial to address these concerns and adapt CWITools to other instruments and it may indeed be so. However, there is always the risk of so-called `unknown unknowns.' Are there assumptions being made that may not apply to all instruments? For example, what if the spaxels in a certain IFS are not spatially adjacent and have gaps between them (say, due to a sparse lenslet array)? The assumption that spaxels are always spatially adjacent is present in the current implementations of PSF modeling and measuring the radial extent of objects. To adjust for this, all such methods would need to be updated to work in world coordinates rather than image coordinates, spatially. The possibility for such unanticipated, fundamental inconsistencies is a major reason for the focused development of CWITools. Extension to other instruments would likely have to be rolled out one at a time, with some period of testing and development required. Having said that, CWITools has intentionally been built in a modular way to allow for this possibility. 

\subsection{Future Development}
The development of this toolkit was motivated in large part by survey work. As discussed briefly in the introduction, the latest generation of IFS instruments on 5-10m class telescopes has enabled surveys of `relatively large' samples of tens of targets. IFS data analysis can be extremely complex and time consuming. If such surveys are to grow in size and remain scientifically productive, data analysis pipelines will be needed. Much of CWITools was developed from scratch because there were no easily available tools to perform the desired tasks. However, it has been developed as an open source package and we extend an open invitation to contributors so that this particular wheel does not need to be reinvented. As such, we encourage anyone in the community who would like to see certain features to contact the authors and get involved. 

\bibliography{cwitools}
\bibliographystyle{aasjournal}
\end{document}